\documentclass[APA,LATO1COL]{WileyNJD-v2}
\index{eq}\usepackage{graphics,graphicx}
\usepackage{amsmath,verbatim,amssymb,epsfig,wrapfig,color,lscape}
\usepackage{longtable}
\usepackage{natbib}
\usepackage{lastpage}
\usepackage{multirow}
\usepackage{url}
\newcommand{\hide}[1]{}

\makeatletter

\usepackage[outerbars,color]{changebar}
\ifx\pdfoutput\undefined
\else\ifnum\pdfoutput>0
  \usepackage{pdfcolmk}
\fi\fi
\cbcolor{black}

\usepackage{tikz}
\usetikzlibrary{bayesnet}
\usetikzlibrary{fit,positioning}

\usepackage{lmodern}

\def\BibTeX{{\rm B\kern-.05em{\sc i\kern-.025em b}\kern-.08em
		T\kern-.1667em\lower.7ex\hbox{E}\kern-.125emX}}

\articletype{Article Type}%

\begin{document}

\title{A Latent Space Accumulator Model for Response Time: Applications to Cognitive Assessment Data}

\author[1,2]{Ick Hoon Jin}
\author[3]{Jonghyun Yun}
\author[1,2]{Hyunjoo Kim}
\author[4]{Minjeong Jeon}

\authormark{Jin \textsc{et al}}

\address[1]{Department of Applied Statistics, Yonsei University. South Korea.}
\address[2]{Department of Statistics and Data Science, Yonsei University. South Korea.}
\address[3]{Institute of Statistical Data Intelligence. USA.}
\address[4]{School of Education and Information Studies, University of California, Los Angeles. USA.}

\corres{Ick Hoon Jin, Department of Applied Statistics, Department of Statistics and Data Science, Yonsei University, Seoul. Republic of Korea. E-Mail: ijin@yonsei.ac.kr.}


\abstract[Summary]{Response time has attracted increased interest in educational and psychological assessment for, e.g., measuring test takers' processing speed, improving the measurement accuracy of ability, and understanding aberrant response behavior. Most models for response time analysis are based on a parametric assumption about the response time distribution. The Cox proportional hazard model has been utilized for response time analysis for the advantages of not requiring a distributional assumption of response time and enabling meaningful interpretations with respect to response processes. In this paper, we present a new version of the proportional hazard model, called a latent space accumulator model, for cognitive assessment data based on accumulators for two competing response outcomes, such as correct vs. incorrect responses. The proposed model extends a previous accumulator model by capturing dependencies between respondents and test items across accumulators in the form of distances in a two-dimensional Euclidean space. A fully Bayesian approach is developed to estimate the proposed model. The utilities of the proposed model are illustrated with two real data examples.}


\maketitle

\section{Introduction}
\label{sec:introduction}

\subsection{Background}\label{sec:intro1}

Response time, i.e., the amount of time test takers spend to give their response to test items, has been a critical concern in educational and psychological research. Researchers measure test takers' response time to evaluate their processing speed, general intelligence,  concentration, and attitude \citep{Fazio:1995, Jansen:1997, Ranger:2013, VanBreukelen:1995}. In addition, researchers examine the relationship between speed and ability because response time can improve the measurement accuracy of ability and help us to understand heterogeneity in response processes that result in aberrant responses, guessing, or different item-solution strategies \citep{DeBoeck:2019, Goldhammer:2015, Kyllonen:2016, Lee:2011}. 

Most approaches to response time are based on a parametric assumption regarding the underlying response time distribution, such as the lognormal distribution \citep{Vanderlinden:2006}, the gamma distribution \citep{Jansen:1997, Maris:1993} and the Weibull distribution \citep{Rouder:2005}. However, the assumed response time distribution is often violated in practice, resulting in biased inferences on the model parameters and respondents \citep{Molenaar:2018, Ranger:2021}. To remedy this issue, semiparametric approaches have been presented to relax stringent distributional assumptions about response time. For instance, \citet{Molenaar:2018} proposed a semiparametric mixture modeling approach to detect differences in responses and response times within subjects. \citet{Liu:2022} presented a semiparametric factor model using a conditional density of the observed response time variable and the latent processing speed variable. \cite{Ranger:2021} applied a factor copula model to connect the marginal response time distributions, approximated via a spline hazard model, to the latent speed of the respondents. 

Another approach adopted to relax the response time distribution assumption is the Cox proportional hazards model \citep{Douglas:1999, Loeys:2014, Kang:2016, RangerK:2012, Ranger:2013, Wang:2013}. The proportional hazards model can be a useful option because it does not require a strong distributional assumption of response time and can also deal with censoring (e.g., test takers do not complete an item in the given amount of time). Furthermore, the model enables a meaningful interpretation of response processes, e.g., how fast a test taker accumulates given information over time until he/she gives a response to the item based on the baseline hazard functions \citep{Ranger:2013}. \citet{Douglas:1999} first proposed a version of the proportional hazards model based on discrete response time in psychometric applications. \citet{Ranger:2012, Ranger:2013} presented a new application of the proportional hazards model for continuous response time from multi-item tests. \citet{Wang:2013} proposed a Cox proportional hazards model with a latent speed covariate within the hierarchical framework \citep{vanderLinden:2007} to jointly model response time and accuracy. \citet{Loeys:2014} proposed a proportional hazards model with crossed random effects to capture heterogeneity due to subjects and items.

Further elaborating the earlier proportional hazards models, \citet{Ranger:2014} presented a Cox model based on competing risks, assuming two accumulators representing different response processes: the one reflecting the progress toward the true response and the other one reflecting the tendency to discontinue working on the item (e.g., non-response). Independent proportional hazards models with different latent traits, different baseline hazard functions, and different response borders represent the two accumulators. This type of Cox model is closely related to the accumulator model presented in the literature \citep{Vickers:1970, vanZandt:2000, Usher:2001, Brown:2005, Brown:2008, Ranger:2014}, in the sense that two hazard functions correspond to two distinct accumulators that describe the response processes between respondents and items. Accumulators are assumed to acquire evidence of each response over time. The one that accumulates sufficient evidence first drives the corresponding response to occur, whereas the other outcome remains unobserved due to mutual exclusiveness. 

\subsection{The Current Paper}\label{sec:intro2}

The accumulator framework adopted by \citet{Ranger:2014} helps us understand the potentially different nature of item solution processes that lead to correct and incorrect responses. We note that \citet{Ranger:2014} specified the hazard function as a function of a baseline hazard function and a person's latent trait ($\theta_p$) and an item parameter ($\beta_i$) where the person and item parameters are multiplied  ($\beta_i \cdot \theta_p)$ in each accumulator. Here, the item parameter $\beta_i$ can be interpreted similarly to item slopes (or factor loadings),  indicating the relationship between the latent trait and the items or the influence of the latent trait on the hazard function for item $i$ \citep{Ranger:2014}. This model assumes that all respondents with the same latent trait level show identical baseline hazard functions, i.e., the rate of accumulation of information over time, for the same items. Similarly, the relationship between the latent trait and the items (i.e., item slopes) is identical for all respondents. This is a strong homogeneity assumption because, despite identical latent trait levels, respondents may show different information accumulation rates because of unobserved cognitive, affective, or attitude differences among respondents in their response processes. In addition, identical items may show different characteristics to different respondents due to unobserved similarities or differences in the properties of the items (e.g., formats, structures, or contents). 

To relax this homogeneity assumption and allow for dependence between respondents and items in the hazard function, we propose a latent space accumulator model based on the proportional hazard model. The main idea is to capture  unobserved dependence between respondents and test items in the hazard functions across the two accumulators in the form of distances between respondents and items in a low-dimensional geometric space. Evaluating the respondent and item configuration of the latent space helps us understand the presence and patterns of heterogeneity between the respondents, between the test items, and between the respondents and the test items. With two empirical examples, we will demonstrate how the dependence structure revealed in the latent space can improve our understanding of respondents and test items in terms of item solution processes. We will use a fully Bayesian approach to estimate the proposed model. 

The method presented in this study is inspired by the latent space item response model \citep[LSIRM;][]{Jeon:2021}, a recent development for item response data that aims to capture and visualize conditional dependencies between responses and items in a low-dimensional geometric space, also called an interaction map. Using response time information, we leverage this idea of LSIRM in the context of evaluating heterogeneity (or dependence) between respondents and between items in an item solution process using response time and accuracy information. 

The remainder of this article is organized as follows. In Section~\ref{sec:model}, we describe the proposed model in detail. In Section~\ref{sec:estimation}, we present the proposed Bayesian inference method. In this section, a simulation-based model assessment is also established. In Section~\ref{sec:applications}, we provide two real data applications of the proposed approach using a computer-based chess game and a mobile language assessment app in Section~\ref{sec:applications}. We close our paper with the conclusions in Section~\ref{sec:conclusions}.

\section{Proposed Model: Latent Space Accumulator Model}
\label{sec:model}

Suppose we have a binary response $x_{ki}$ from the respondent $k$ ($k = 1, \cdots, n$) to the item $i$ ($i = 1, \cdots, p$), where $x_{ki}$ is associated with the response time $t_{ki}$, which is the observed time that the respondent $k$ spent answering the item $i$, regardless of the outcome of the item response (e.g., correct or incorrect). Based on the response time information, our goal is to capture patterns of dependence among the respondents and the items. 

In this setting, a notable feature is the two response outcomes are mutually exclusive (e.g., correct vs. incorrect responses), and the dependence structure of response times is likely to be heterogeneous across the two response outcomes. To characterize the heterogeneous dependence structures for the two mutually exclusive response outcomes, we assume that the respondents and the items have positions in a low-dimensional Euclidean space \citep{Hoff:2002, Handcock:2007, Krivitsky:2009, Raftery:2012, Friel:2016, Jeon:2021}, where the distances between the respondents and the items indicate dependence between the pairs in the latent space. Given that it is a type of accumulator model equipped with a latent space, we refer to our proposed model as a latent space accumulator model. 

\subsection{Model Structure}

\subsubsection{Hazard Function}

In the proposed latent space accumulator model, a hazard function is specified for the response outcome $c$: 
\begin{equation}\label{eq:hazard_lsam1}
  h_{kic}(t) = \lambda_{ic}(t) \exp\Big(\theta_{kc} + c \cdot ||\mathbf{z}_{k} - \mathbf{w}_{i}||\Big),
\end{equation}
where $c \in \{-1,1\}$ represents incorrect and correct responses, respectively, $\theta_{kc}$ represents the latent trait of the respondent $k$ for the response $c$, $\lambda_{ic}(t)$ represents an unspecified baseline hazard function of the item $i$ for the response $c$, $\mathbf{z}_{k} \in \mathbb{R}^{d}$ and $\mathbf{w}_{i} \in \mathbb{R}^{d}$ are embedded latent positions of the respondent $k$ and the item $i$ in the $d$-dimensional Euclidean latent space, and $||\cdot||$ represents the Euclidean norm. The latent trait of the respondent $k$, $\theta_{kc}$, constitutes the propensity of the respondent $k$ toward the response $c$ in the hazard function. 

The latent distance in the hazard function aims to account for heterogeneous interactions between the respondents and the items in terms of response times for two response outcomes. As distance $||{\bf z}_k - {\bf w}_i||$ increases, the hazard function for the positive response $c= 1$ increases, leading to a decrease in response times. In contrast, a decrease in the hazard function for the negative response $c=-1$ leads to an increase in response times. As a result, a large distance between respondents and items in a latent space indicates a large difference in response times between the two possible response outcomes. On the other hand, a short distance between the respondents and the items indicates little difference in response times between the two response outcomes. An estimated latent space configuration helps us to understand the differences in response times between the two response outcomes between respondents and test items. 

The baseline hazard function is specific to the response $c$ and depends on the tendency of the latent trait of the item $i$ toward the response $c$. We assume a piecewise constant baseline hazard function \citep{Ibrahim:2001}, which provides flexible learning of the unspecified dependence between the response and the response times. We place $J+1$ points $0 = s_0 < s_1 < \cdots < s_J < \infty$ that are used for the piecewise exponential approximation of the baseline hazard function:
\begin{equation}\label{eq:hazard_piecewise1}
  \lambda_{ic}(t) = \lambda_{ic,j} \quad \text{if} \quad s_{j-1} \le t < s_{j},
\end{equation}
for $j = 1, 2, \cdots, J$. This approximation casts the cumulative baseline hazard function as
\begin{align*}\label{eq:hazard_piecewise2}
  \Lambda_{ic}(t) & = (t - s_{j-1})\lambda_{ic,j} + \sum_{m=0}^{j-1} (s_{m} - s_{m-1})\lambda_{ic,m},
\end{align*}
for $s_{j-1} \leq t < s_j$. A baseline hazard $\lambda_{ic,j}$ can be interpreted as the propensity of the item $i$ toward response $c$ on the hazard function in a given time interval $s_{j-1} \leq t < s_j$.

The interpretation of the key model parameters is given as follows:

\begin{itemize}
    \item $\lambda_{ic,j}$ in the baseline hazard function represents the accumulation rate for a specific response $c$ that occurs to a particular item $i$ in a given time interval $(s_{j-1}, s_j)$. With the accumulation rate difference between two possible responses in a given time interval $(s_{j-1}, s_j)$ for an item $i$, $\Delta\lambda_{i,j} = \lambda_{i(-1),j} - \lambda_{i(1),j}$, we can identify a response time difference as well as a more probable outcome for item $i$ within the time interval $j$. A small $\Delta \lambda_{i,j}$ indicates that the item $i$ has similar response times between both possible outcomes in the time interval $j$, whereas a large $\Delta \lambda_{i,j}$ indicates that the item $i$ has distinct response times for each response in the time interval $j$, implying that item $i$ has a specific response that is more likely to occur in the time interval $j$. When $\Delta\lambda_{i,j}$ is positive, item $i$ is more likely to produce a negative response in the time interval $j$, implying that the respondent is likely to get the item wrong in an assessment setting. Note that $\lambda_{ic,j}$ can be interpreted as a latent property parameter for the item $i$ to the response $c$ in a given time interval $(s_{j-1}, s_j)$.

    \item $\theta_{kc}$ in the response-specific hazard function is a latent trait that represents the accumulation rate for the response outcome $c$ that occurs to a particular respondent $k$. Similarly to $\lambda_{ic,j}$, we can calculate the accumulation rate differences in $\theta_{kc}$ between the two possible responses, $\Delta \hat{\theta}_{k} = \hat{\theta}_{k(-1)} - \hat{\theta}_{k(1)}$. These differences help us identify which response is more likely to occur for the respondent $k$. In other words, the differences can notify whether a respondent is likely to get correct or incorrect in an  assessment setting. A small $|\Delta \theta_{k}|$ indicates that the respondent $k$ shows similar response times between the two possible responses. On the other hand, a large $|\Delta \theta_{k}|$ implies that the respondent $k$ has different response times between the two response outcomes, indicating that either response can occur for the respondent $k$ in overall periods.  

    \item An estimated latent space for response times is referred to as an interaction map in this paper, following \citet{Jeon:2021}. The interaction map offers additional insights on the accumulation rates between respondent $k$ and the item $i$ for the two possible outcomes $c$. 
    A shorter distance between the respondent $k$ and item $i$ implies that the accumulation rates between the  two possible outcomes are similar, indicating negligible differences in response times between the respondent $k$ and the item $i$. In contrast, a larger distance between the respondent $k$ and the item $i$ indicates that the accumulation rates between the two possible outcomes are not negligible, implying that there is a more probable response for respondent $k$ and item $i$. 
\end{itemize}

Note that in the hazard function we specified in Equation (\ref{eq:hazard_lsam1}),  the item parameters $\beta_i$ do not appear, unlike  \citet{Ranger:2014}'s specification. As mentioned in Section \ref{sec:intro2},  the item parameters $\beta_i$ indicate the relationship between the latent traits of the respondents and the test items. Since $\beta_i$ is multiplied by the latent trait term for persons, the item parameters can capture interactions between persons and items to some degree  \citep{Jeon:2021}. By dropping $\beta_i$ from the proposed model, the distance term $||{\bf z}_k - {\bf w}_i||$ is set to capture all possible interactions or relationships between respondents and items in the corresponding accumulator. Furthermore, due to the triangle inequality of distances, the distance term additionally captures interactions between respondents and respondents, as well as between items and items \citep{Jeon:2021}. In other words, our specification with the distance term $||{\bf z}_k - {\bf w}_i||$ captures interactions more generally in item response time data than in \citet{Ranger:2014}'s model with the item parameters $\beta_i$. 

\subsubsection{Overall Survival Function}

Since the response outcomes are mutually exclusive, we can write the overall survival function under the framework of the competing risk model \citep{Fine:1999, Lau:2009, Andersen:2012, Ranger:2014}. Then, the overall survival function $S_{ki}(t) = P(T_{ki} > t)$ can be expressed as
\begin{equation}\label{eq:hazard_lsam2}
S_{ki}(t)  = \exp\bigg(-\int_{0}^{t} \sum\nolimits_{c\in \{-1,1\}} h_{kic}(s) ds\bigg).
\end{equation}
The joint density of response times and outcomes can be given as
\begin{equation}\label{eq:hazard_lsam3}
f\Big(T_{ki} = t, X_{ki} = a\Big) = h_{kia}(t) \exp\bigg\{-\sum\nolimits_{c \in \{-1, 1\}} \exp\big\{\theta_{kc} + c \cdot||\mathbf{z}_{k} - \mathbf{w}_{i}||\big\} \Lambda_{ic}(t)\bigg\}.
\end{equation}

Let $\mathbf{\Theta} = \{\theta_{kc}\}_{1 \le k \le n, ~c = \{-1,1\}}$ denote a set of parameters for the latent traits of the respondents and $\mathbf{\Lambda} = \{\lambda_{ic,j}\}_{1 \le i \le p, ~1 \le j \le J, ~c=\{-1,1\}}$ denote a set of baseline hazards with respect to the items. Denote $\mathbf{Z}= (\mathbf{z}_{1}, \cdots, \mathbf{z}_{n})$ and $\mathbf{W} = (\mathbf{w}_{1},\cdots,\mathbf{w}_{p})$ as the configuration of latent embeddings. Conditional on the latent traits and embeddings, we assume that $x_{ki}$ and $t_{ki}$ are independent. Then, the likelihood of our Bayesian latent space accumulator model is given by
\begin{equation*}
\begin{split}
    f\Big(\mathbf{X}, \mathbf{T} &\mid \mathbf{\Theta}, \mathbf{\Lambda}, \mathbf{Z}, \mathbf{W}\Big) 
    = \prod_{k=1}^n \prod_{i=1}^p \prod_{j=1}^J  \prod\nolimits_{c=\{-1, 1\}} \left\{\lambda_{ic,j} \exp\Big(\theta_{kc} + c \cdot ||\mathbf{z}_k - \mathbf{w}_i||\Big) \right\}^{\delta_{ki,j} \nu_{kic}}\\
    &\times \exp\left[-\delta_{ki,j}\Big\{ (t_{ki} - s_{j-1})\lambda_{ic,j} + \sum_{m=0}^{j-1} (s_{m} - s_{m-1})\lambda_{ic,m} \Big\} \exp\Big(\theta_{kc} + c \cdot ||\mathbf{z}_k - \mathbf{w}_i||\Big) \right],
\end{split}
\end{equation*}
where
\[ 
    \delta_{ki,j} = \left\{\begin{array}{ll} 1 & \mbox{if } s_{j-1} \leq t_{ki} < s_{j},\\ 0 & \mbox{otherwise,} \end{array}\right. \quad \mbox{and} \quad \nu_{kic} = \left\{\begin{array}{ll} 1 & \mbox{if } x_{ki} = c, \\ 0 & \mbox{otherwise.} \end{array}\right.
\]

\subsubsection{Cumulative incidence function}
We will examine the cumulative incidence function (CIF) that quantifies the probability that one event occurs before time $t$ and before the occurrence of the competing event \citep{Austin:2016}. Due to the presence of competing responses, the CIF is an attractive analysis tool for the proposed approach as it helps us investigate varying degrees of respondent-item interactions over time. Specifically, the CIF of the response $c$ for nodes $k$ and $i$ is defined as: 
\[ \mbox{CIF}_{kic}(t) = Pr\Big(T_{ki} \leq t, Y_{ki} = c\Big). \] 
CIF can be thought of as the probability that respondent $k$ has a specific outcome before time $t$. The CIFs of respondents located closely in an interaction map should be similar across responses $c$. When respondent $k$ and item $i$ are far apart in an interaction map, a relatively steep-and-tall CIF for $c = 1$ is expected. When respondent $k$ and item $i$ are closely located to each other, on the other hand, a relatively gradual-and-short CIF for $c = 1$ is expected. 

\subsection{Advantages}

\begin{itemize}
\item {\bf Practical advantages}: A unique advantage of the proposed model, compared to other conventional models for response time, is its ability to capture and represent the inherent interactions (or dependence) among respondents and items in response time and visualize the dependence in a low-dimensional space. This geometric representation offers the detection of unobserved characteristics related to items and responses in terms of response time, providing insights into the heterogeneity in the accumulation rates across respondents and items. Further, the information from the proposed model, such as the accumulation rates and the cumulative incidence functions, could help improve our understanding of the differences in the item solution processes between correct and incorrect responses.

\item {\bf Weaker assumptions}: The proposed model assumes that the item response times are independent conditional on the positions of respondents and items in the latent space and the response and item attributes. This conditional independence assumption is weaker than the conditional independence of other conventional  models, which requires that response times are independent conditional on the respondent and item attributes. This weaker conditional independence assumption allows for respondent-item interactions in response time, and therefore, the latent space model can account for local dependence among response times arising from a variety of sources.
\end{itemize}

\section{Estimation and Model Fit Assessment}
\label{sec:estimation}

\subsection{Bayesian Estimation}
\label{sec:model-estimation}
We propose a fully Bayesian approach for estimating the proposed latent space accumulator model. Bayesian inference is preferable to maximum likelihood due to the under-identification of the latent embedding. For each $k$, $i$, $c$, $j$, we assume independent priors as follows:
\begin{align*}
  \pi\left(\lambda_{ic,j}\right) & \sim \mbox{Gamma}\left(0.5 \tilde \lambda_{ic,j}, 0.5\right), \quad \pi\left(\theta_{kc} | \sigma^{2}\right) \sim \mbox{N}\left(0, \sigma^{2}\right),
  \quad \pi\left(\sigma^{2}\right) \sim \mbox{Inv-Gamma}\left(a_{\sigma}, b_{\sigma}\right),\\
  \pi\left(\mathbf{z}_{k}\right) &\sim \mbox{MVN}_{d}\left(\mathbf{0}, \gamma^{2}\mathbf{I}_{d}\right), \quad \pi\left(\mathbf{w}_{i}\right) \sim \mbox{MVN}_{d}\left(\mathbf{0}, \gamma^2 \mathbf{I}_{d}\right), \quad \mbox{and} \quad \pi(\log \gamma) \sim \mbox{N}\left(\mu_{\gamma}, \tau_{\gamma}^{2}\right),
\end{align*}
where $\mbox{Gamma}(a,b)$ denotes the gamma distribution with mean $a/b$ and variance $a/b^{2}$, $\mbox{MVN}_{d}$ denotes a $d$-dimensional normal distribution, and $\mathbf{I}_{d}$ is a \(d \times d\) identity matrix. We assign a vague prior for $\lambda_{ic,j}$ by setting $\tilde{\lambda}_{j}=J / \left\{s_J \left(J-j+0.5\right)\right\}$ following from \citet{Jin:2014JASA}. Other hyperparameters are chosen as $a_{\sigma}=0.0001, b_{\sigma}=0.0001, \mu_{\gamma}=0, \text { and } \tau_{\gamma}^{}=2$.

Based on our experience, the inference of \(\mathbf{\Theta}\) is highly sensitive to the variance parameter of $\theta_{kc}$, \(\sigma^2\). Additionally, the configuration of latent embeddings depends on the scale parameter \(\gamma\) of the latent space. To avoid selecting sub-optimal tuning parameters, we introduce a layer of hyper priors, \(\pi\left(\sigma^{2}\right)\) and \(\pi(\log \gamma)\), to learn optimal values of these parameters from data. We choose hyperparameters such that priors are minimally informative to facilitate flexible Bayesian learning.

The Gibbs sampling is employed to obtain posterior samples from
\begin{equation*} \begin{split}
    \pi\Big(\boldsymbol{\Theta}, \boldsymbol{\Lambda}, \mathbf{Z}, \mathbf{W}, \gamma, \sigma^{2} | \mathbf{Y}, \mathbf{T}\Big) &\propto f\Big(\mathbf{Y}, \mathbf{T} \mid \boldsymbol{\Theta}, \boldsymbol{\Lambda}, \mathbf{Z}, \mathbf{W}\Big)\\
    &\times \pi\Big(\boldsymbol{\Theta} \mid \sigma^2\Big) \pi\Big(\boldsymbol{\Lambda}\Big) \pi\Big(\mathbf{Z} \mid \gamma\Big) \pi\Big(\mathbf{W} \mid \gamma\Big) \pi\Big(\sigma^2\Big) \pi\Big(\gamma\Big).
\end{split} \end{equation*}
The random-walk Metropolis-Hastings (MH) algorithm is used to draw samples from the full conditionals for $\mathbf{\Theta}$, $\mathbf{Z}$, $\mathbf{W}$, and $\gamma$. The posterior samples for $\mathbf{\Lambda}$ and $\sigma^{2}$ are directly sampled from their full conditional distributions. One iteration of the MCMC sampler can be described in the supplementary material.

We use (multivariate) Gaussian distributions centered at the current values of the parameters and the latent embeddings as symmetric proposal distributions, with diagonal variance-covariance matrices. For the proposal distribution of $\gamma$, $g_{\gamma}(\cdot \to \cdot)$, we use a log-normal distribution. Variances of the proposal distributions are tuned to achieve an acceptance ratio close to 0.3. To detect non-convergence of the MCMC algorithm, we use trace plots along with the Gelman-Rubin diagnostic \citep{Gelman:1992}. The MCMC algorithm was written in {\tt R} \citep{r_core_team_r_2020} and {\tt C++} with {\tt Stan} math library \citep{Carpenter:17}. The code, along with example data sets, is found in \url{https://github.com/Jonghyun-Yun/LSA}.

Each outcome-specific hazard function is invariant to translations, reflections, and rotations of the latent positions of respondents and items because the hazard function depends on the positions through the distances, and the distances are invariant under the aforementioned transformations. As a consequence, the likelihood function is invariant under these transformations like latent space models of network data \citep{Hoff:2002}. Such identifiability issues can be resolved by post-processing the MCMC output with Procrustes matching \citep{Gower:1975}.

\subsection{Model Fit Assessment}

To assess model fit, data are simulated from the posterior predictive distribution $p\big(\mathbf{\tilde Y}, \mathbf{\tilde T} \mid \mathbf{Y}, \mathbf{T}\big)$, where $\mathbf{\tilde Y} = \{\tilde Y_{ki}\}$ and $\mathbf{\tilde T} = \{\tilde T_{ki}\}$ denote the generated outcome and response times for respondent $k$ and item $i$. The Cox model with piecewise baseline functions leads to a monotone decreasing survival function, and its analytic inverse exists \citep{walke_example_2010}. Thus, we use the probability inverse transformation based on the overall survival function to generate the response times.

Let $g_{ki,j}$ denote the overall hazard function at $j$-th segment and is given as follows:
\[ g_{ki,j} = \sum_{c=\{-1,1\}} \lambda_{ic,j}\exp\Big\{\theta_{kc} + c \cdot ||\mathbf{z}_k - \mathbf{w}_i||\Big\}. \]
Then, the inverse transformation of the overall survival function is given as
\begin{equation}\label{eq:inverse1}
t=s_{j}+\frac{1}{g_{ki,j}}\left[- \log \Big\{S_{ki}(t)\Big\} - \sum_{m=1}^{j} g_{ki,m} \Big(s_{m}-s_{m-1}\Big) \right],
\end{equation}
if the survival function is bounded by cumulative overall hazards
\begin{equation}\label{eq:survival}
 \sum_{m=1}^{j} g_{ki,m}\Big(s_{m}-s_{m-1}\Big) < - \log \Big\{S_{ki}(t)\Big\} \leq  \sum_{m=1}^{j+1} g_{ki,m}\Big(s_{m}-s_{m-1}\Big),
\end{equation}
for some \(j \in \{1,\ldots, J-1\}\); otherwise the inverse transformation becomes
\begin{equation}\label{eq:inverse2}
t = \left\{\begin{array}{ll} \log \Big\{S_{ki}(t)\Big\} \Big/ g_{ki,1}, & \mbox{if}~ -\log\Big\{S_{ki}(t)\Big\} <  g_{ki,1}s_{1},\\
s_J, & \mbox{if}~ \sum_{m=1}^{J} g_{ki,m}\Big(s_{m}-s_{m-1}\Big) < -\log \Big\{S_{ki}(t)\Big\}.
\end{array}\right.
\end{equation}
Given the response times $\tilde T_{ki}$, the conditional distribution of the outcome is
\begin{equation}\label{eq:connection_type}
P(\tilde Y_{ki} = c \mid \tilde{T_{ki}} = t) = \frac{h_{kic}(t)}{\sum_{c \in \{-1,1\}} h_{kic}(t)}
\end{equation}
for each $k$ and $i$.

Let $L$ denote the total number of MCMC iterations. Algorithm 1 in Section 2 of the Supplementary Material elaborates on how to generate posterior predictive samples of $\mathbf{\tilde {\bf Y}}$ and $\mathbf{\tilde {\bf T}}$. The posterior predictive p-values, or Bayesian p-values, are used to assess the model fit to response times. The p-values are calculated for each $k$ and $i$ to quantify the discrepancy between the simulated and observed response times as follows:
\begin{equation}
\label{eq:ppp}
\mbox{P}_{B_{ki}} = \frac{1}{L}\sum_{l=1}^{L}I(\tilde{T}_{ki}^{(l)} \geq t_{ki}).
\end{equation}
The p-value outside $[0.05,0.95]$ is thought to be evidence of a model misfit. We use classification performance metrics to assess the model fit to the response outcomes. Let
\begin{equation}\label{eq:pred_p}
    {p_{kic}^{(l)} \equiv P(Y_{ki} = c | T_{ki} = \tilde{T}_{ki}^{(l)})},
\end{equation}
which can serve as the prediction probability of \(Y_{ki} = c\) connection type based on $l$-th iteration of MCMC samples. To see if these probabilities are close to the observed truth \(Y_{ki}\), we calculate the log-loss as follows:
\begin{equation}\label{eq:logloss}
\mbox{Log-loss}_{i}^{(l)} =  - \frac{1}{N} \sum_{k=1}^{N}I(Y_{ki} = c) \log p_{kic}^{(l)},
\end{equation}
for each item $i$ and respondent $l$. For an additional metric, we estimate the receiver operating characteristic (ROC) curves for each item $i$ and respondent $l$ by treating $Y_{ki} = 1$ as positive outcome. Then, the area under the curve (AUC) is obtained to summarize the overall assessment of our model.

\section{Real Data Applications}
\label{sec:applications}

In this section, we apply the proposed  latent space accumulator model to two real data examples. Both applications are based on computerized assessments where a group of test-takers responds to a set of test items, where the two competing outcomes are correct and incorrect responses, while response times indicate how long it took for the test-takers to give responses to the test items.  
By applying the MCMC algorithm described in Section~\ref{sec:model-estimation}, we generate the posterior distributions of the model parameters for each example. Three independent MCMC chains are initiated with random starting values. Each chain consisted of 20,000 iterations, with the first halves discarded as a burn-in. From the remaining samples, we retrieved every $10$-th draw (thinning interval) to construct posterior samples. 
In interpreting the estimated results, we focus on accumulation rate differences ($\Delta\lambda_{i,j}$), interaction maps, and cumulative incidence functions (CIFs). 

\subsection{Amsterdam Chess Test}\label{sec:act}
\subsubsection{Data and Estimation}

The first dataset is obtained from a sub-test of the Amsterdam Chess Test (ACT) designed to measure players' chess playing proficiency \citep{vanderMass:2016}. The dataset consists of 40 items divided into three sets based on the required chess skills: 20 tactical items, 10 positional items, and 10 endgame items with increasing difficulty in each set. Tactical items measure a player's ability to calculate a sequence of moves to obtain tangible results, whereas positional items measure how proficient a player is at improving his/her own position by obtaining a superior position while decreasing the opponent’s tactical potential. The endgame items mainly measure the player's skills necessary in situations where few pieces are left. Players were asked to find an optimal move for each item within a 30-second time limit. The information on response accuracy and response time was recorded per item for each player. 

\subsubsection{Analysis Results}
\label{sec:analysis-results}

\begin{figure}[htbp]
    \centering
    \includegraphics[width=15cm, height=20cm]{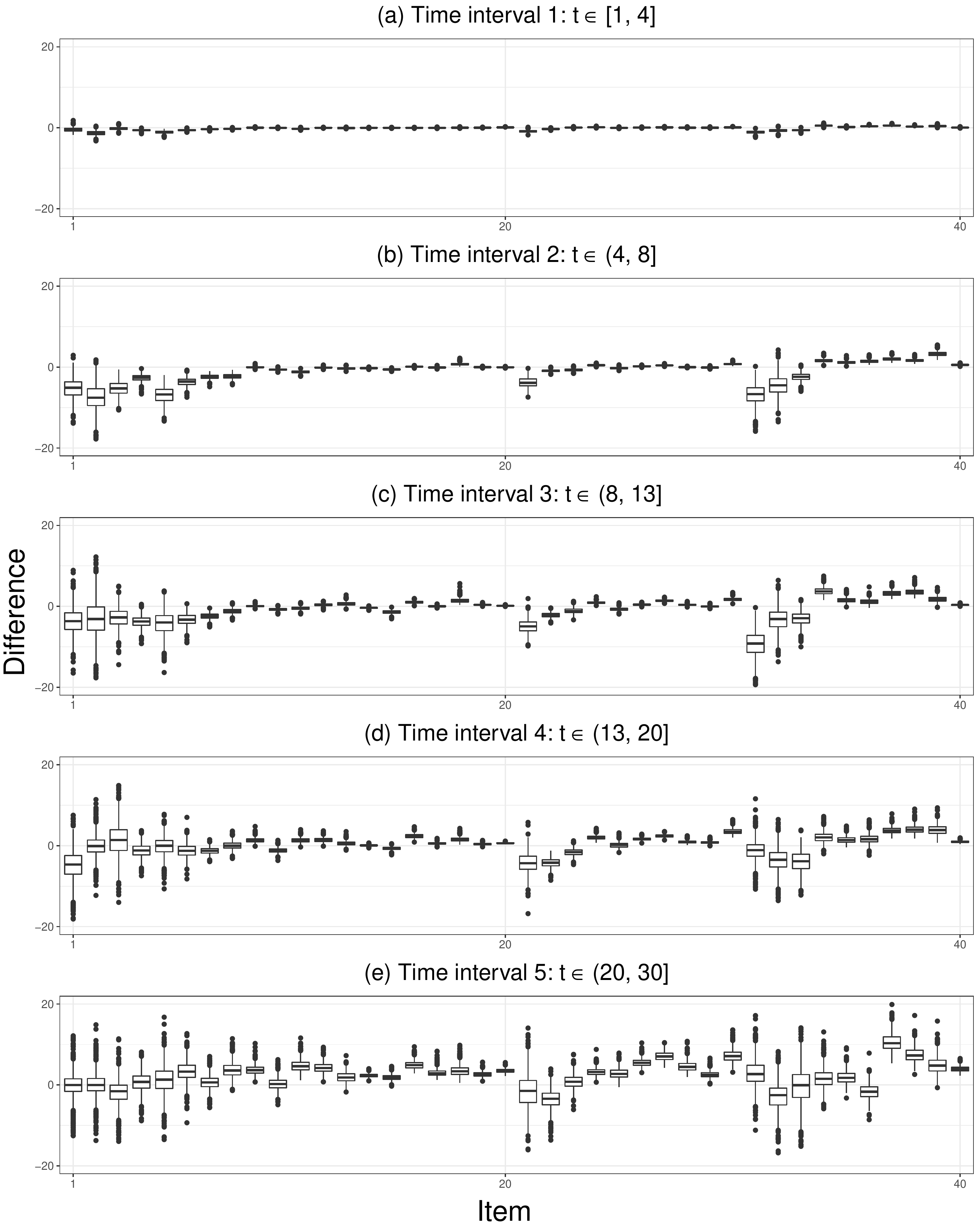}
     \caption{\label{fig:lambda_chess}%
    The posterior distribution of \(\Delta\lambda_{i,j} = \lambda_{i(-1),j} - \lambda_{i(1),j}\). The response time \(t\) is divided into five time intervals \(( j=1,2,\ldots,5 )\), and \(\Delta\lambda_{i,j} \) for each item \(j\) are calculated for the time
    intervals.}
    \label{fig:my_label}
\end{figure}

\paragraph*{Accumulation Rate Differences \(\Delta\lambda_{i,j}\)} 

The posterior distributions of $\Delta\lambda_{i,j} = \lambda_{i(-1)j} - \lambda_{i(1)j}$ in the 40 test items are presented in Figure~\ref{fig:lambda_chess}. The values of \(\Delta\lambda_{i,j}\) represent accumulation rate differences in the baseline hazard functions between incorrect and correct responses for each time interval. Response times were divided into five subintervals using the sample quantiles as cutoff points: 1 to 4 seconds for the first interval; 4 to 8 seconds for the second interval; 8 to 13 seconds for the third interval; 13 to 20 seconds for the fourth interval; and 20 to 30 seconds for the fifth interval.


As described in Section~\ref{sec:model}, the analysis of $\Delta\lambda_{i,j}$ across time interval $j$ reveals the individual item's distribution of response accuracy and times. For example, items 1-3 show negative mean values of $\Delta \lambda_{i,j}$ in earlier time intervals. These are fairly easy items where roughly 75\% of the respondents correctly answered them before $t=8$ seconds. On the other hand, Items 37-39 show $\Delta \lambda_{i,j} \approx 0$ in earlier time intervals (near $t=1$), while the distributions are shifted upward to the end of the time intervals. These are difficult items with low accuracy (10 - 30 \%), and response times are evenly distributed over the five-time intervals. 

\paragraph*{Interaction Map} 

\begin{figure}[htbp]
    \centering
    \begin{tabular}{cc}
    (a) & (b) \\
    \includegraphics[width=0.475\textwidth]{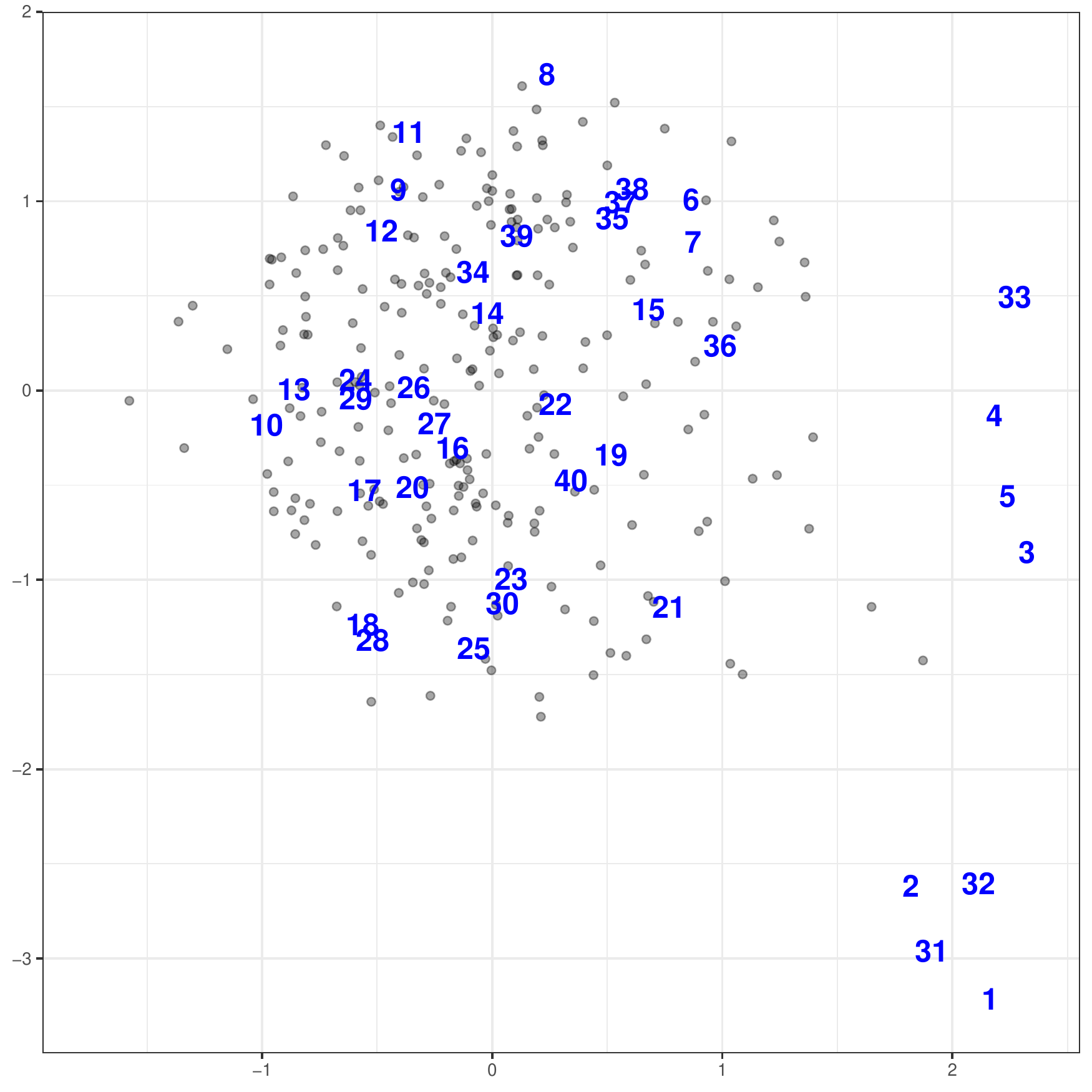} & 
    \includegraphics[width=0.475\textwidth]{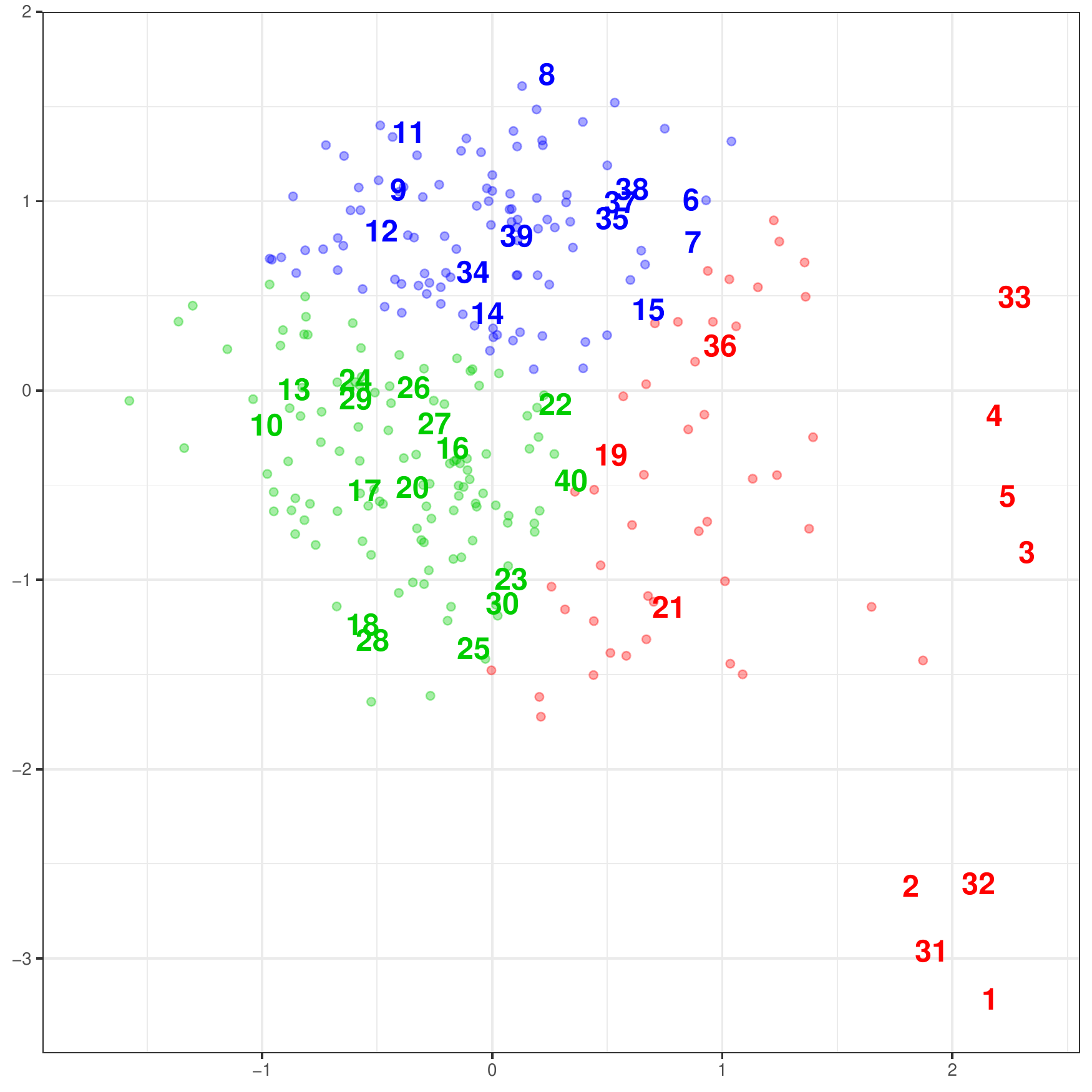} \\
    (c) & (d) \\ 
    \includegraphics[width=0.45\textwidth]{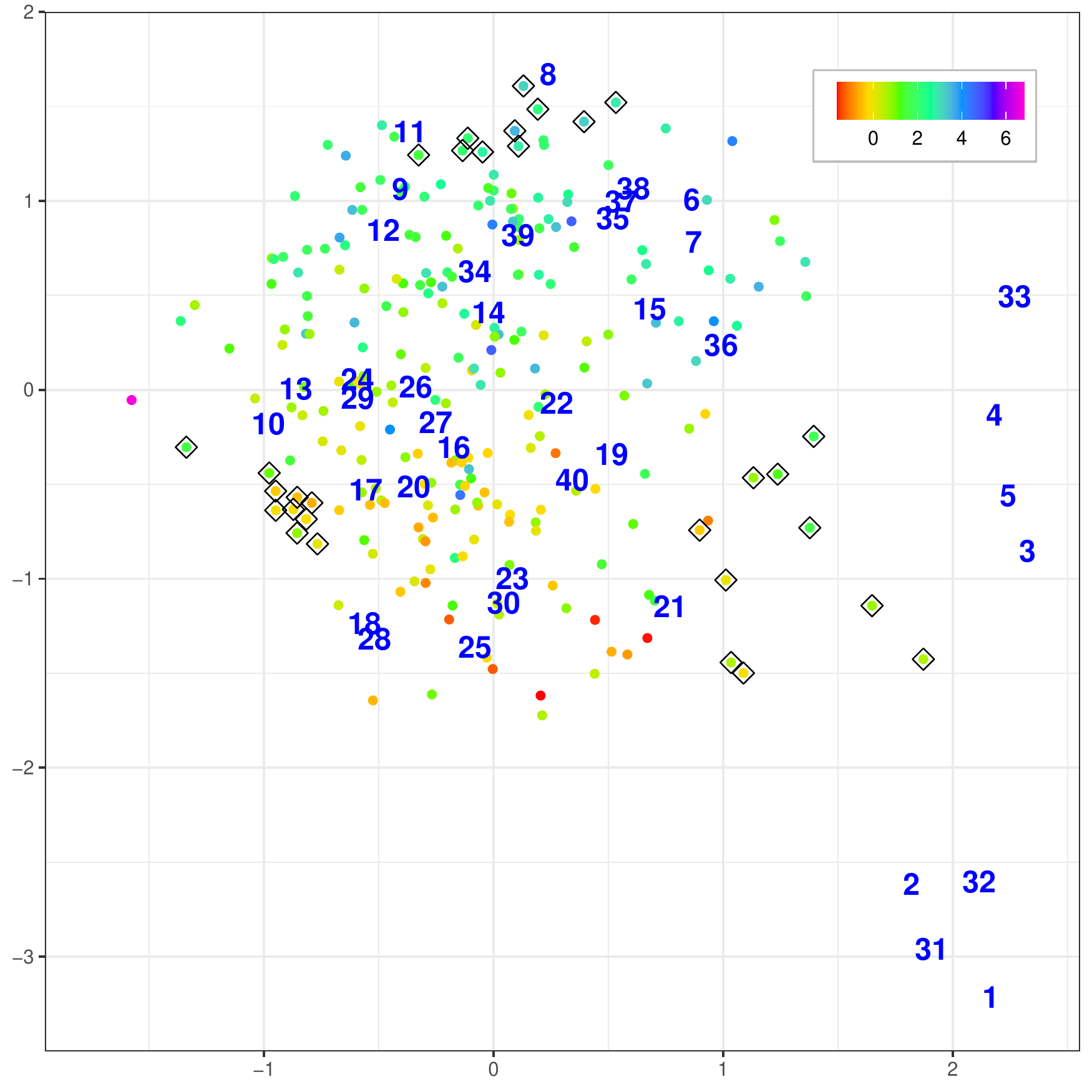} & 
    \includegraphics[width=0.45\textwidth]{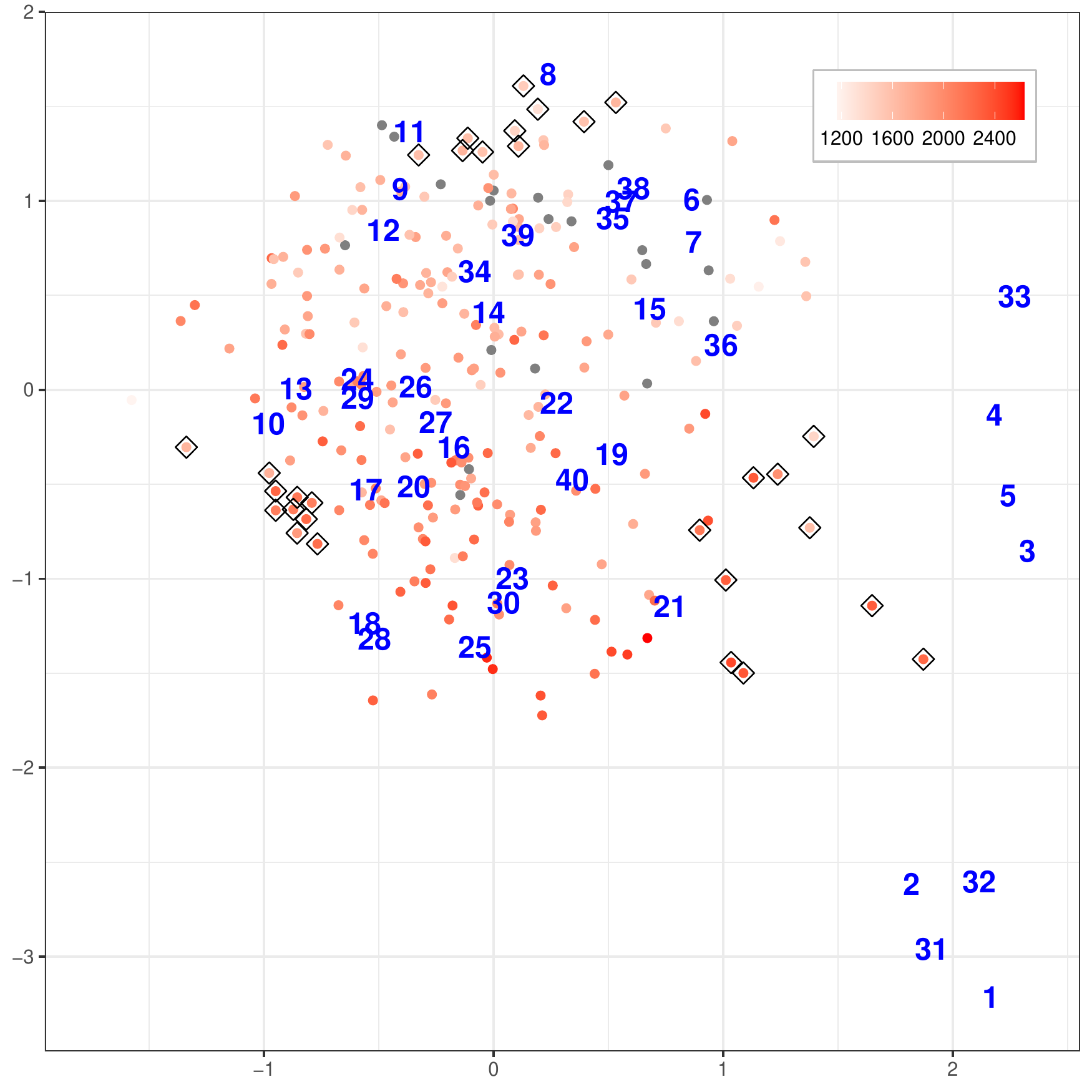} \\
    \end{tabular}
    \caption{
    \label{fig:int_chess}
    Interaction maps of latent embeddings estimated by the Bayesian latent space accumulator model for ACT data. (a) Interaction Map. Dots represent respondents, and numbers represent items. (b) Interaction map with the spectral co-clustering result. Groups of items and respondents are marked using red, green, and blue. (c) Interaction map with the posterior mean differences. Posterior mean difference, ($\hat{\Delta\theta}_{k} = \hat{\theta}_{k(-1)} - \hat{\theta}_{k(1)}$), are converted to a red scale. (d) Interaction map with the ELO rating. ELO ratings are converted into the red scale, with white being the lowest and red being the highest.
    }
\end{figure}

Figure~\ref{fig:int_chess}(a) is the estimated interaction map for this data set. Each subfigure consists of dots representing respondents and numbers representing items. The distance between a respondent $k$ and an item $i$ indicates the difference in accumulation rates between the two possible outcomes. A larger distance indicates a significant difference, suggesting a strong tendency toward one type of outcome. This indicates that the respondent was able to solve the items relatively quickly and exhibited high certainty toward the correct answers. Consequently, a high ratio of observed correct responses is observed. This pattern is typically observed when respondents encounter relatively easy items. For example, items 1-5 and 31-33 are located far from the majority of respondents, indicating that these items are generally easy items (i.e., high accuracy) with short response times for most respondents. 

On the contrary, a shorter distance between the respondent $k$ and the item $i$ implies a negligible difference in the accumulation rates between two possible outcomes, meaning that there is a minor difference in response times between the correct and incorrect outcomes when the respondents solve the items. This means that respondents experience uncertainty in their answers, leading to longer response times. This is reflected in a low ratio of observed correct responses and prolonged decision-making.

To interpret the interaction map more clearly, individual respondents are grouped with the items they are close to. To this purpose, we applied the spectral co-clustering \citep{Dhillon:2001} based on a radial basis function of $||\mathbf{z}_{k} - \mathbf{w}_{i}||$. We determined that $K = 3$ was the optimal number of clusters based on the elbow method. The result of the co-clustering is presented in Figure~\ref{fig:int_chess}(b), with the cluster membership represented in red, green, and blue. When the shorter the distance between item and respondent latent positions, the more likely they are to be grouped in co-clustering, respondents are expected to respond long with low accuracy to items in their cluster membership. 

Upon careful examination of the distances between items and respondents in Figure \ref{fig:int_chess} (b), we can discern the relationship between items and respondents in terms of accuracy and response time. For example, the respondents at the bottom of the green cluster are notably distant from items 35, 37, and 38, meaning a higher likelihood of solving these items correctly with a relatively short response time. However, these respondents were closely located to items 18 and 28, suggesting a higher probability of answering incorrect responses with longer response times. Note that all five items (items 18, 28, 35, 37, and 38) exhibit low accuracy rates, ranging from approximately 10\% - 40\%. This result demonstrates that our approach can differentiate subtle individual differences in both the success probability and the response times of these difficulty items. 

In addition, we compute the differences in the accumulation rates $\theta_{kc}$ between the two response outcomes, denoted as $\Delta\hat{\theta}_{k} = \hat{\theta}_{k(-1)} - \hat{\theta}_{k(1)}$, where $\hat{\theta}_{kc}$ represents the posterior mean of $\theta_{kc}$. In Figure \ref{fig:int_chess}(c), we visualize the respondents overlaid with $\Delta\hat{\theta}_{k}$ using a continuous colored scale. As the red color gets darker, it indicates that $\Delta\hat{\theta}_{k}$ approaches zero, suggesting a minimal difference in response time between the two possible outcomes for the respondent $k$. In contrast, a high $\Delta\hat{\theta}_{k}$ implies that the respondent $k$ tends to provide accurate answers to items. Therefore, there is a negative correlation between $\Delta\hat{\theta}_{k}$ and the accuracy of the response. In the ACT example, the correlation between $\Delta\hat{\theta}_{k}$ and overall accuracy is -0.952. 

We find that respondents near the boundaries of the space between the red and green clusters tend to have small values of $\Delta\hat{\theta}_{k}$, indicating overall high accumulation rates towards correct responses. On the contrary, in the blue cluster and the top of the red cluster, respondents tend to show lighter red dots, suggesting that they have overall high accumulation rates toward incorrect responses.

Lastly, we overlay individual respondents with their ELO ratings, an external measure representing the chess playing competency of chess players \citep{vanderMass:2016}. In Figure~\ref{fig:int_chess}(d), respondents' ELO ratings are converted into a continuous red scale; the lighter the red scale, the lower the ELO rating. For instance, a white dot indicates a respondent with a very low ELO rating. According to the figure, respondents in the green group have higher ELO scores, whereas those in the blue group have lower ELO scores. The red group appears to include those respondents with low and high ELO ratings, but those in the upper section have lower ELO ratings than those in the lower section, similar to the blue and green groups. 

\paragraph*{Cumulative Incidence Functions (CIFs)} 

We examine the CIFs for selected items and respondents from each of the three clusters. Specifically, we chose one item from each cluster: Item 11 (blue), Item 28 (green), and Item 3 (red). We chose ten respondents from each cluster and marked them in hollow circles in Figures~\ref{fig:int_chess}(c) and \ref{fig:int_chess}(d). 

\begin{figure}[htbp]
\centering
\begin{tabular}{ccc}
\multicolumn{3}{c}{Correct Responses} \\
(a) Item 11 (blue) & (b) Item 28 (green) & (c) Item 3 (red) \\
\includegraphics[page=1,width=0.3\textwidth]{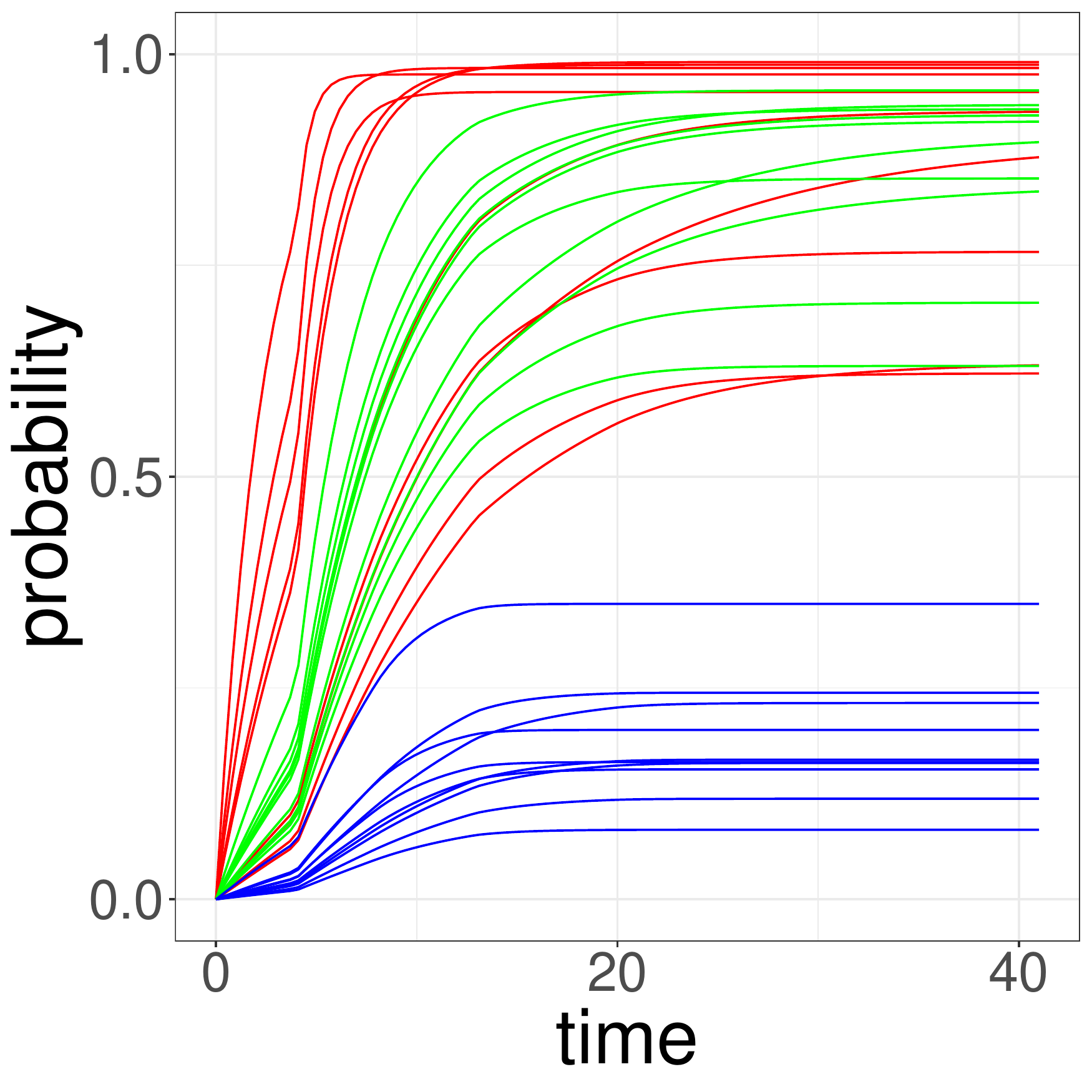} &
\includegraphics[page=3,width=0.3\textwidth]{figure_sub/cif_chessb.pdf} &
\includegraphics[page=5,width=0.3\textwidth]{figure_sub/cif_chessb.pdf} \\
\multicolumn{3}{c}{Incorrect Responses}\\
(d) Item 11 & (e) Item 28 & (f) Item 3\\
\includegraphics[page=2,width=0.3\textwidth]{figure_sub/cif_chessb.pdf} &
\includegraphics[page=4,width=0.3\textwidth]{figure_sub/cif_chessb.pdf} &
\includegraphics[page=6,width=0.3\textwidth]{figure_sub/cif_chessb.pdf} \\
\end{tabular}
\caption{Cumulative incidence functions (CIFs) of selected respondents and items for ACT data. We select Item 11 from the blue cluster, Item 28 for the green, and Item 3 from the red. Each of the 10 respondents is chosen nearby their cluster centers, and marked by their cluster membership colors. We present correct response CIFs in the top panel and incorrect response CIFs in the bottom.
}
\label{fig:cif_chess}
\end{figure}

Figure~\ref{fig:cif_chess} shows the CIFs for the selected items when the responses are correct (top row) and incorrect (bottom row). The CIFs for the ten selected respondents are drawn in each plot, where the line colors indicate their cluster membership. The CIF of a respondent for correct and incorrect responses to the same item shows the opposite patterns, as shown in Figure~\ref{fig:cif_chess}. Therefore, our analysis will focus on the CIFs for correct responses to conserve space. 

First, the peak height of the CIF corresponds to the respondent's probability of giving a correct response. For example, respondents in blue generally have lower peaks than the other respondents for Item 11, indicating that the blue group has a lower likelihood of getting Item 11 correct than the other groups. Respondents in green show lower peaks in the CIFs for item 28 compared to the two other items. This means that the group had a lower likelihood of giving Item 28 the correct answer compared to the other two items.

Second, how fast a CIF reaches its peak indicates how long it would take to give a correct response. For example, for Item 11, the CIFs of some respondents in the red group reach their peaks rapidly, while other respondents in the same red group reach their peaks much more slowly. This means that there was some degree of individual differences in the red group in terms of how long they would need to spend to give a correct response to Item 11. Recall that respondents and items were co-clustered when they are close in the interaction map, the CIFs tend to reach their peaks slowly when the respondents and items are in the same group (e.g., respondents in blue for Item 11).

Lastly, we find that $\Delta \hat{\theta}_{k}$ ($> 1.5$) are positive in the blue group, while $\Delta \hat{\theta}_{k}$ are negative and similar in the red and green groups. Note that despite the similarity in the global statistic $\Delta \hat{\theta}_{k}$, the red and green groups are clearly differentiated in terms of CIFs, as shown in Figure \ref{fig:cif_chess}. This suggests that examining CIFs provides additional insights to evaluate individual differences in the response processes.

\paragraph*{Model Fit: Posterior Predictive Checks} 

\begin{figure}[htbp]
\centering
\includegraphics[page=1,width=0.32\columnwidth]{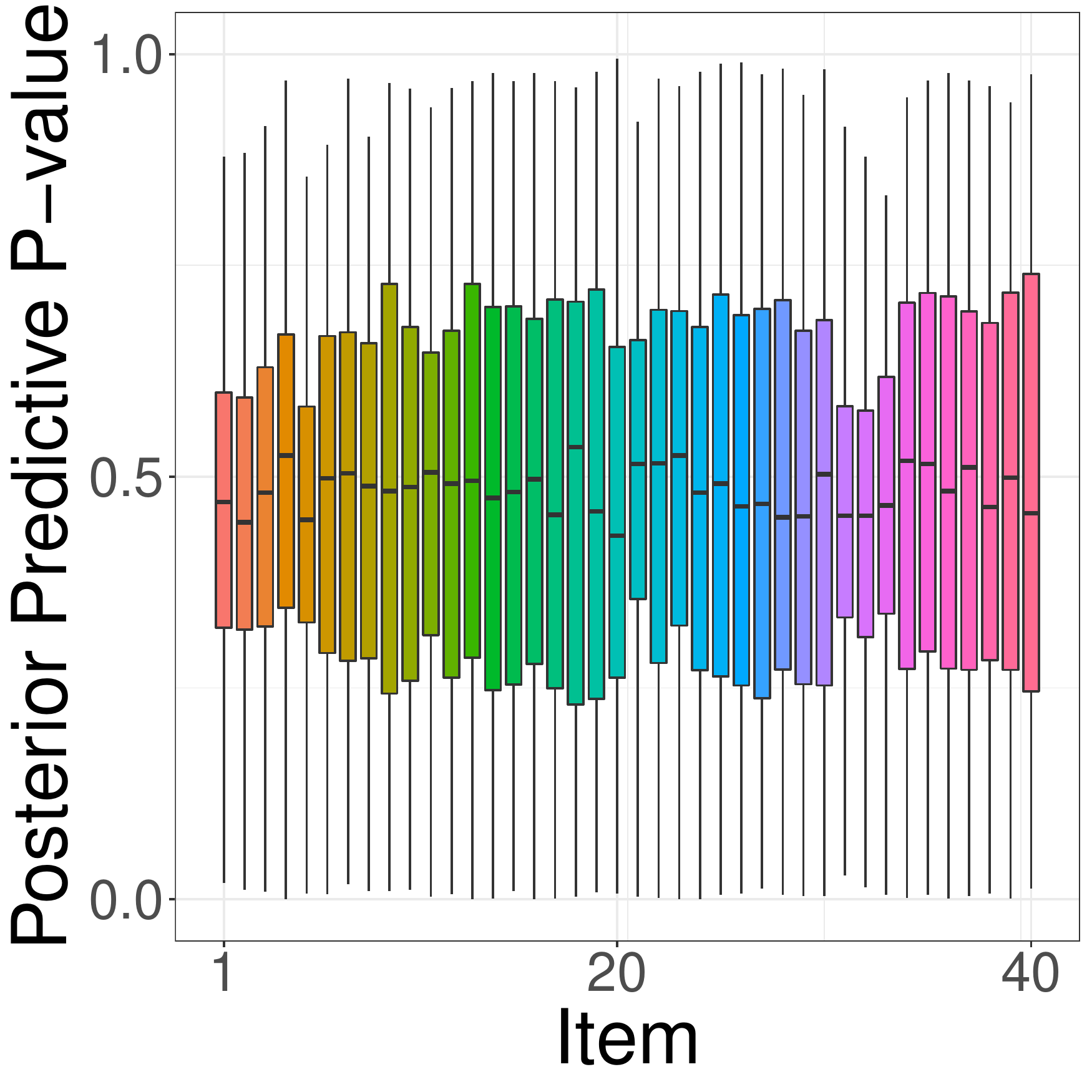}
\includegraphics[page=1,width=0.32\columnwidth]{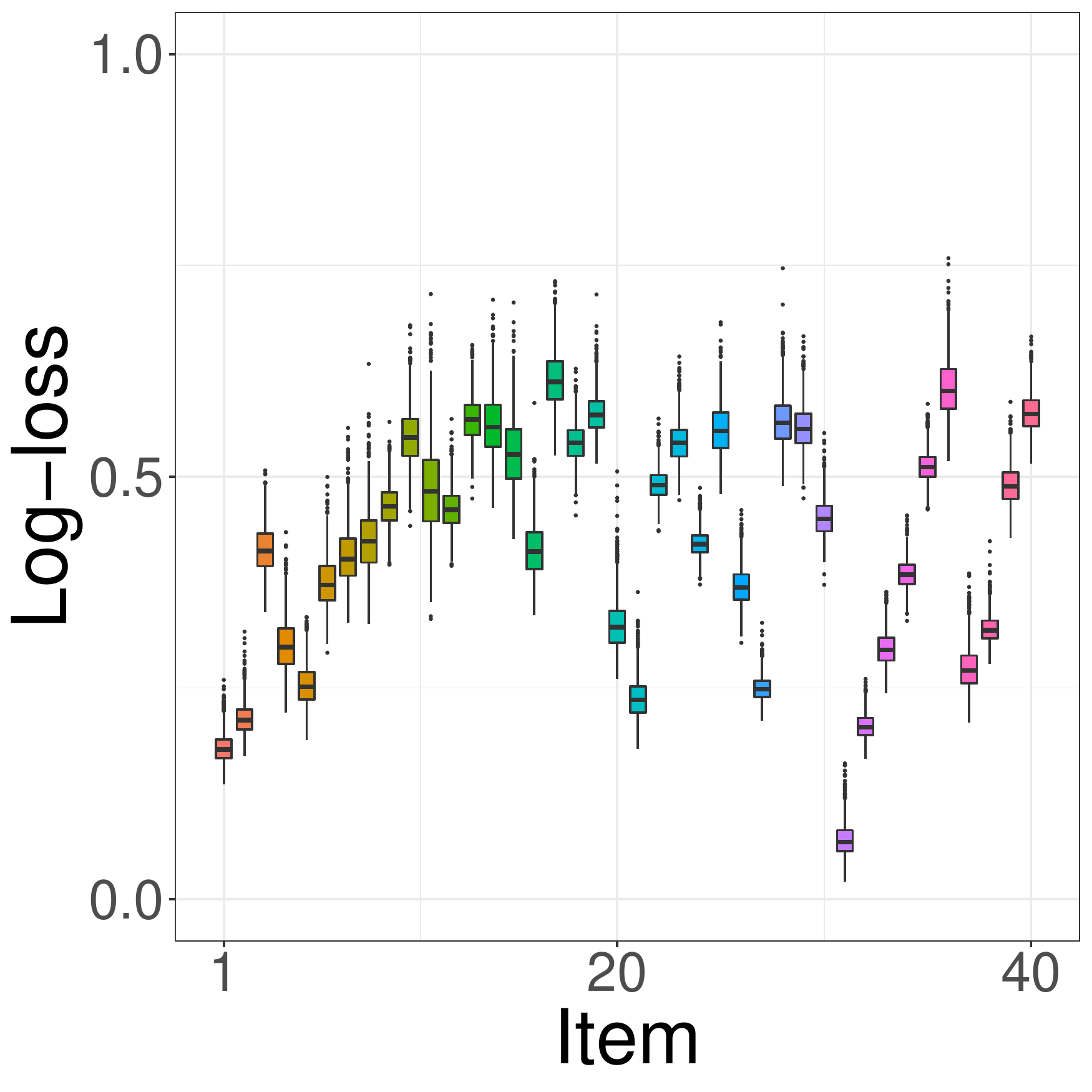}
\includegraphics[page=2,width=0.32\columnwidth]{figure_sub/chessB_pp_type.pdf}
\caption{
\label{fig:chessB_pp}
Model fit assessment for ACT data. Posterior predictive samples are generated to calculate posterior predictive p-values (left panel), the log-loss (middle panel), and AUC (right panel) for each item. }
\end{figure}

We generate posterior predictive samples of size 1,000 using Algorithm 1 in Section 2 of the Supplementary Material. The posterior predictive p-values \eqref{eq:ppp}, log-loss values \eqref{eq:logloss}, and AUCs are presented in \ref{fig:chessB_pp}. The posterior predictive p-values range between 0.05 and 0.95 for all items. AUCs range between 0.7 and 0.84 for most items. Log-loss values range from 0.2 to 0.6 for most items. We find that the log-loss values are relatively large for difficult items. This may result from evenly distributed response times for difficult items. Overall, these results suggest a reasonable fit of the proposed model for the ACT data.

\subsection{App-based Language Assessment}\label{sec:duolingo}

\subsubsection{Data and Estimation}

The second data example is obtained from the Duolingo mobile language learning app Duolingo. Data were released for ``The 2018 Duolingo Shared Task on Second Language Acquisition Modeling''\footnote{\href{http://sharedtask.duolingo.com/2018.html}{http://sharedtask.duolingo.com/2018.html}}. The data set includes responses to a set of test items given to individual app users. For analysis, we chose translation items, where users can translate a short sentence written in Spanish into English fully or by arranging given English words. For example, an item asks to translate a Spanish sentence `\textit{Yo estoy bien.}' into English: `\textit{I am fine.}', or to arrange three given words, ``I'', ``am'', and ``fine'', into the correct order. Listening items were not included due to the large volume of missingness. There are two additional features to be discussed: (1) not all items are given to all test users. In other words, the number and types of items are different per user. For the sake of simplicity, we selected the users who responded to a common set of items; (2) there can be multiple responses to the same items given by the same users as the assessment system exposes identical items to users multiple times. We included the responses obtained at the first exposure if multiple responses appeared. As a result, a total of 151 users responded to 18 items that were included for data analysis. 

This second data example from the app-based assessment is different from the ACT data set analyzed in Section \ref{sec:act} in three ways. First, in the Duolingo data, the test items are similar to each other in terms of format, structure, content, and difficulty levels. This means that the items are less distinguishable among themselves in this example than the items in the ACT data. Second, no time limit is set for each item and for the entire test. This means that users can spend on an item as long as they want, resulting in substantial variation in response times across items and respondents in this data set. Lastly, students' motivation levels vary to a large degree in this app-based, non-conventional kind of assessment. A lack of motivation may be responsible for large within- and between-person variations in response accuracy and times.

\begin{figure}[htbp]
\centering
\includegraphics[width=15cm, height=20cm]{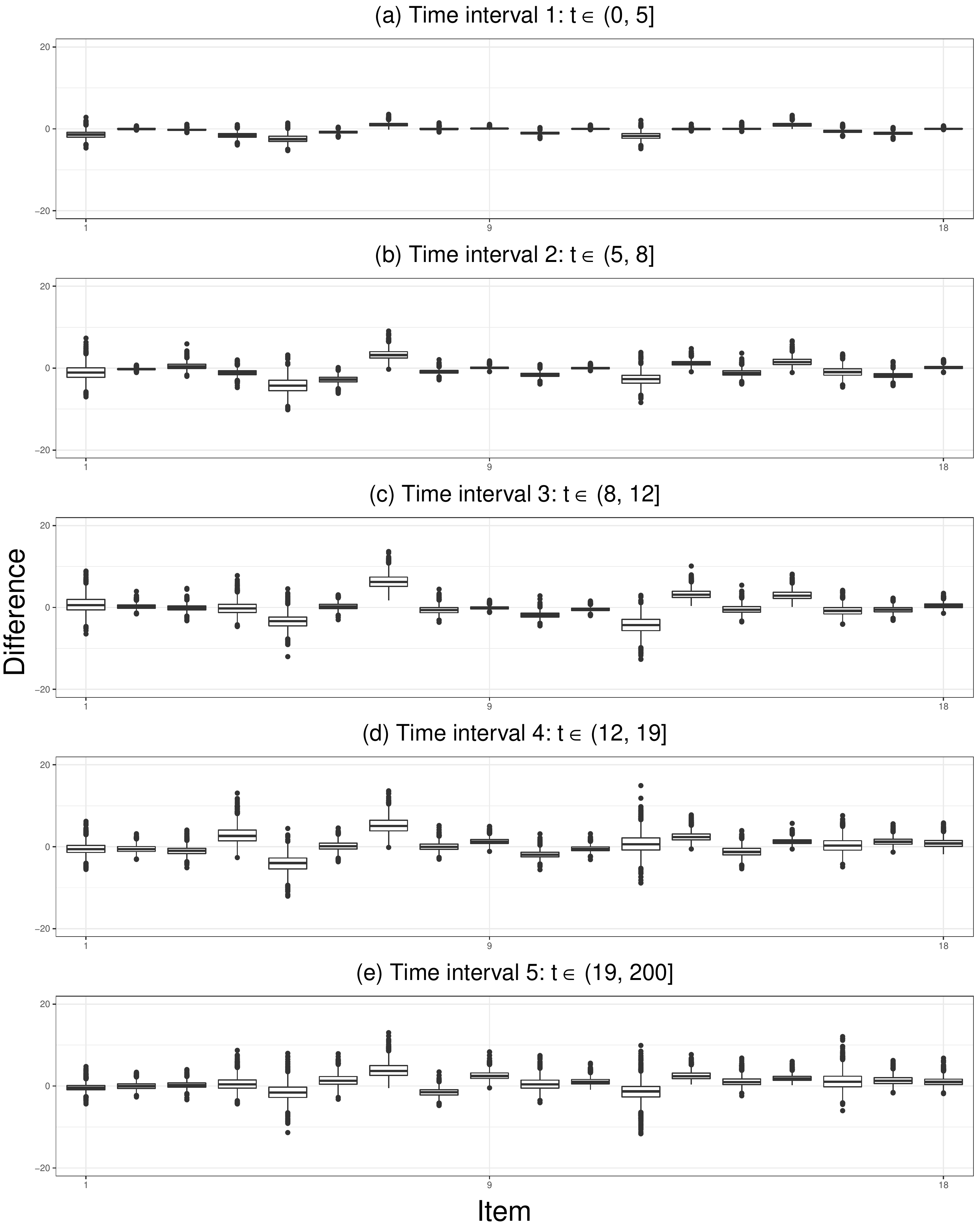}
\caption{\label{fig:duo_lambda}
The posterior distribution of \(\Delta\lambda_{i,j} = \lambda_{i(-1),j} - \lambda_{i(1),j}\). The response time \(t\) is divided into five time intervals \(( j=1,2,\ldots,5 )\), and \(\Delta\lambda_{i,j} \) for each item \(j\) are calculated for the time intervals.}
\end{figure}

\subsubsection{Analysis Results}


\paragraph*{Accumulation Rate Differences \(\Delta\lambda_{i,j}\)} 

Figure~\ref{fig:duo_lambda} summarizes the posterior distribution of $\Delta{\lambda}_{i,j}$. Response time was divided into five sub-intervals using the sample quantiles as cut-off points: 0 to 5 seconds for the first interval; 5 to 8 seconds for the second interval; 8 to 12 seconds for the third interval; 12 to 19 seconds for the fourth interval; and 19 to 200 seconds for the fifth interval. Response time over 200 seconds was censored. Figure~\ref{fig:duo_lambda} shows that $\Delta\lambda_{i,j}$ are around 0 and rarely change over time intervals for all items. This indicates that the accumulation rates between correct and incorrect responses are similar and do not change much over time. This is different from the ACT data case, where meaningful differences were observed in accumulation rate across the time intervals and test items.

\paragraph*{Interaction Map} 

\begin{figure}[htbp]
\centering
\begin{tabular}{cc}
(a) & (b) \\
\includegraphics[width=0.45\textwidth]{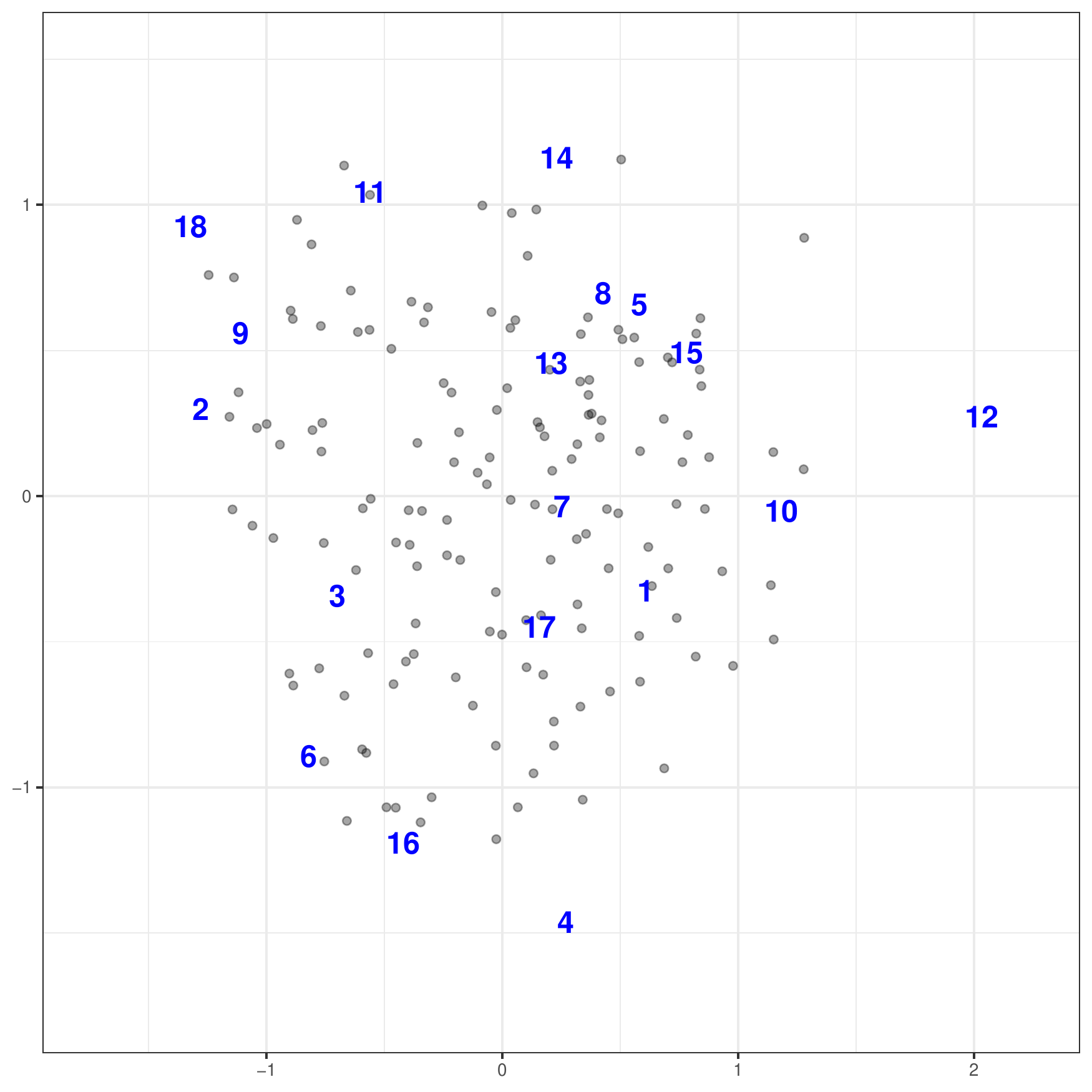} & 
\includegraphics[width=0.45\textwidth]{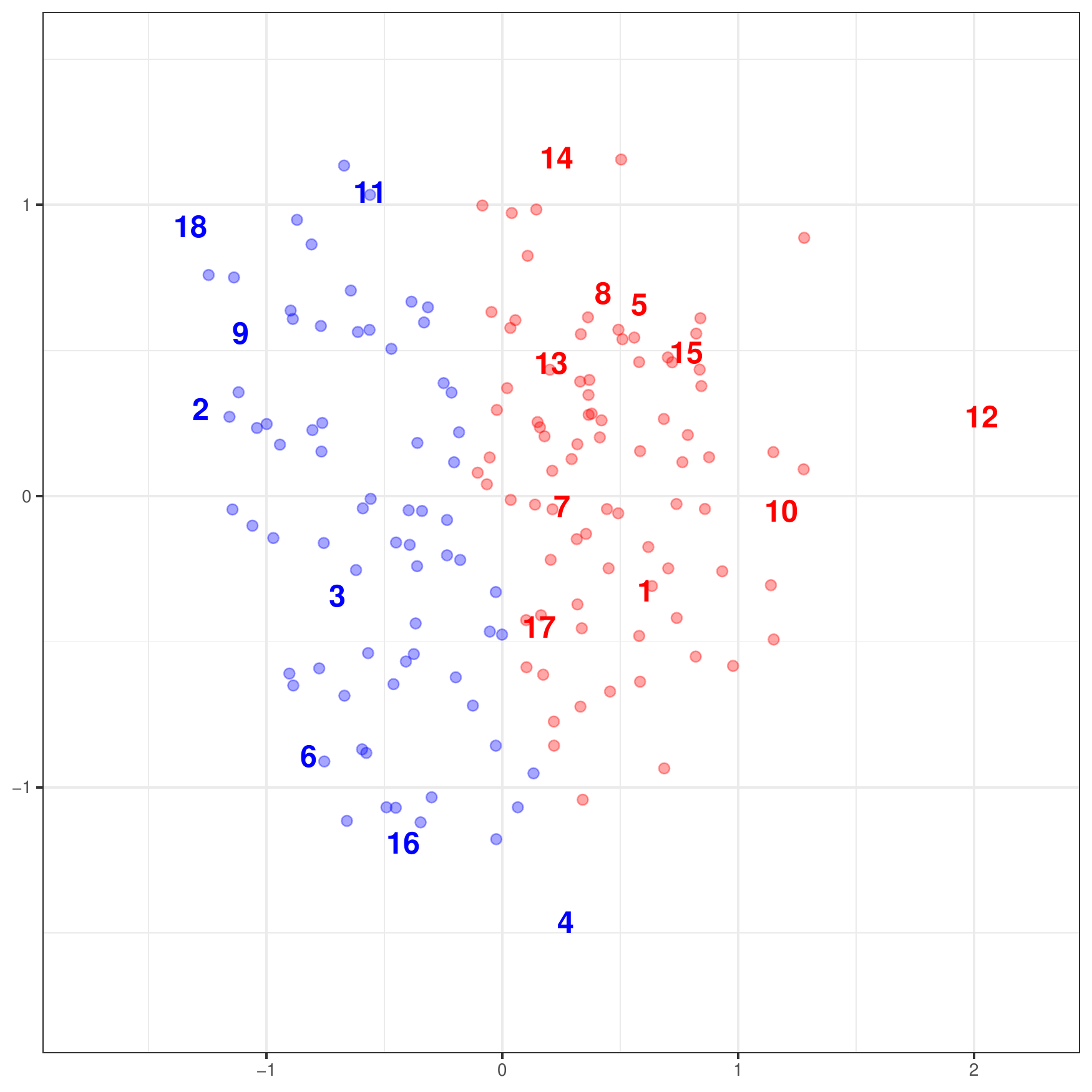} \\
(c) & (d) \\
\includegraphics[width=0.45\columnwidth]{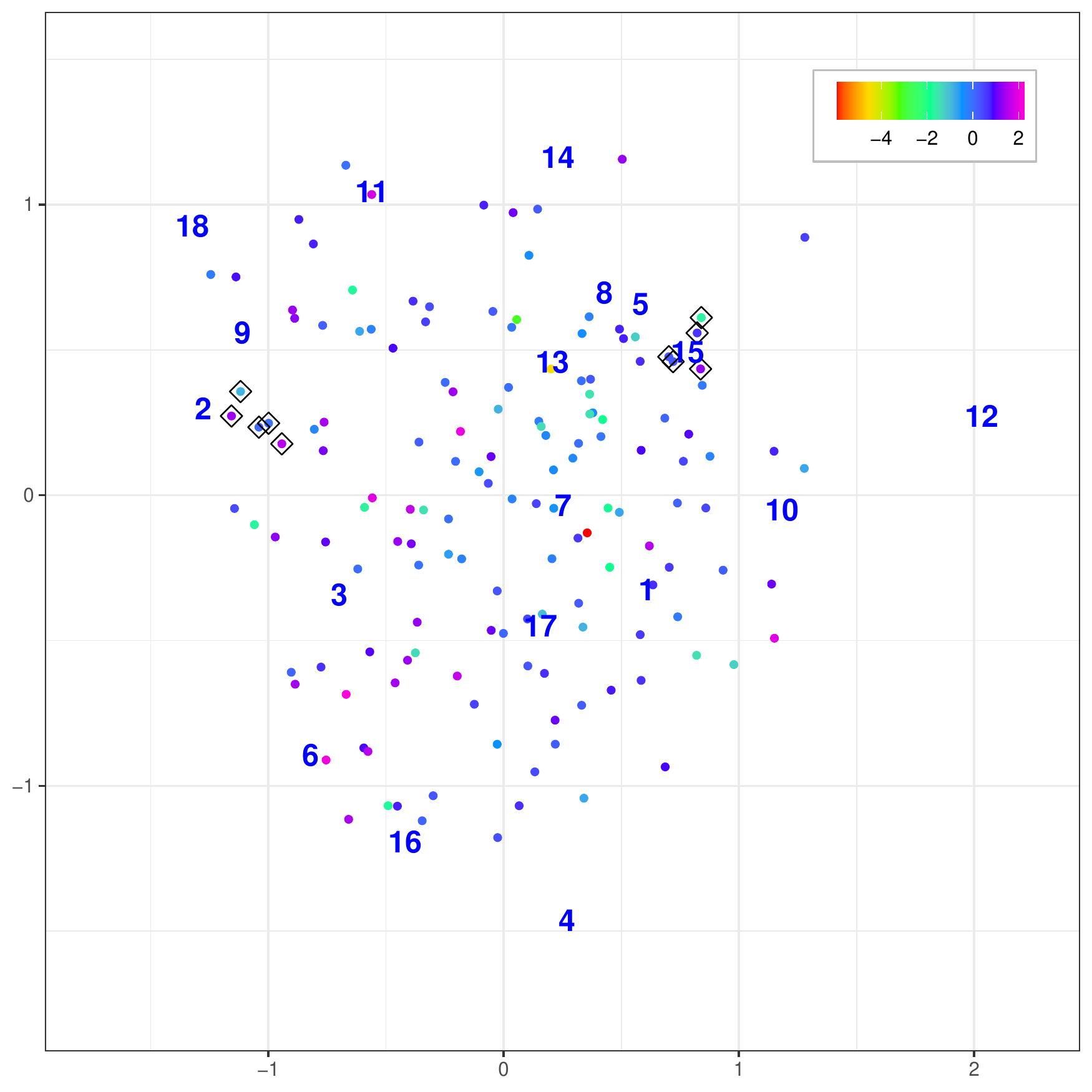} &
\includegraphics[width=0.45\columnwidth]{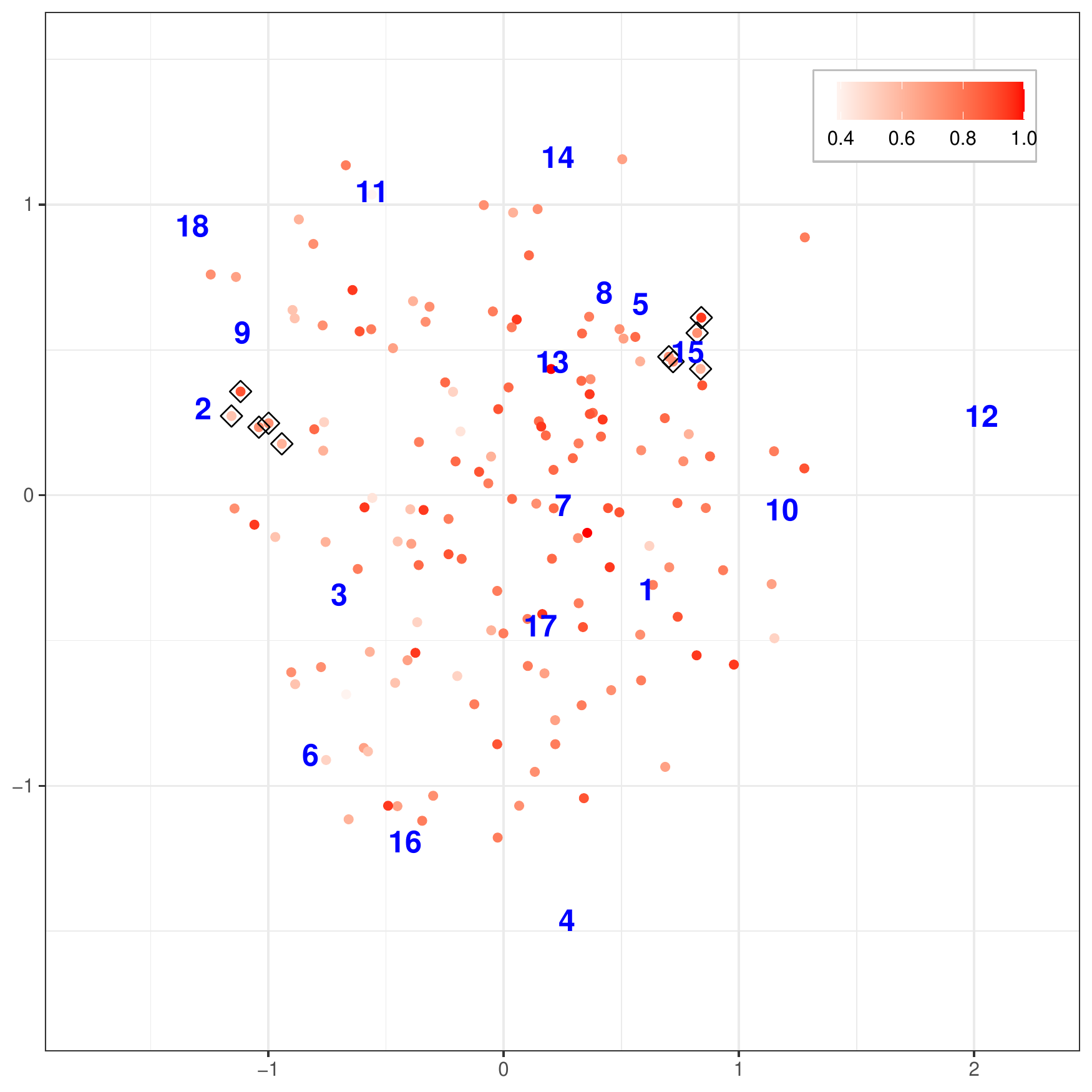}
\end{tabular}

\caption{Interaction maps of latent embeddings estimated by the Bayesian latent space accumulator model for Duolingo data. (a) Interaction Map. Dots represent respondents and numbers represent items. (b) Interaction map with the spectral co-clustering result. Items and respondent groups are marked in red and blue. (c) Interaction map with the posterior mean differences. Posterior mean difference, ($\hat{\Delta\theta}_{k} = \hat{\theta}_{k(-1)} - \hat{\theta}_{k(1)}$), are converted to a red scale. (d) Interaction map with the ELO rating. ELO ratings are converted to the red scale with white being the lowest and red being the highest.}
\label{fig:int_duolingo}%
\end{figure}

The estimated interaction maps for the Duolingo data are presented in Figure~\ref{fig:int_duolingo}(a). As shown in Section \ref{sec:analysis-results}, interaction maps can differentiate subtle differences in the correctness of responses as well as the length of response times. For example, items 6, 8, 12, and 14 are located on the outskirts of the map, meaning that they were solved correctly by most users in a relatively short time compared with other items. We found that these are relatively simple questions that only involve one or two words.

As in the ACT example, spectral co-clustering is applied to identify sets of item-person pairs that are close to each other, and the result is shown in Figure \ref{fig:int_duolingo}(b). Using the elbow method, $K = 2$ is chosen as the optimal number of clusters in this dataset. The cluster membership is represented in red and blue colors.

In Figure \ref{fig:int_duolingo}(c), we overlay the respondents with $\Delta{\hat\theta}_{k}$ as in the ACT example (Section~\ref{sec:analysis-results}). The red cluster includes near-zero or negative $\Delta{\hat\theta}_{k}$, implying that the respondents' accumulation rates were relatively higher when they gave a correct response than an incorrect response. The blue cluster includes respondents with positive $\Delta{\hat\theta}_{k}$ values, meaning that their accumulation rates were relatively higher when they gave incorrect responses than correct responses. 
Lastly, we overlay respondents with average response accuracy in Figure \ref{fig:int_duolingo}(d). The respondents from the red group show generally higher response accuracy than the respondents in the blue group, consistent with the findings based on $\Delta{\hat\theta}_{k}$.

\paragraph*{Cumulative Incidence Functions (CIFs)} 

We then examined CIFs for selected items and respondents from each cluster. We chose one item from each cluster: Item 2 (blue) and Item 15 (red). We selected five respondents from each cluster and marked them in hollow circles in Figures \ref{fig:int_duolingo}(c) and \ref{fig:int_duolingo}(d). 

\begin{figure}[htbp]
\centering
\begin{tabular}{cc}
\multicolumn{2}{c}{Correct Responses} \\
(a) Item 2 (blue)  & (b) Item 15 (red)  \\
\includegraphics[page=1,width=0.45\textwidth]{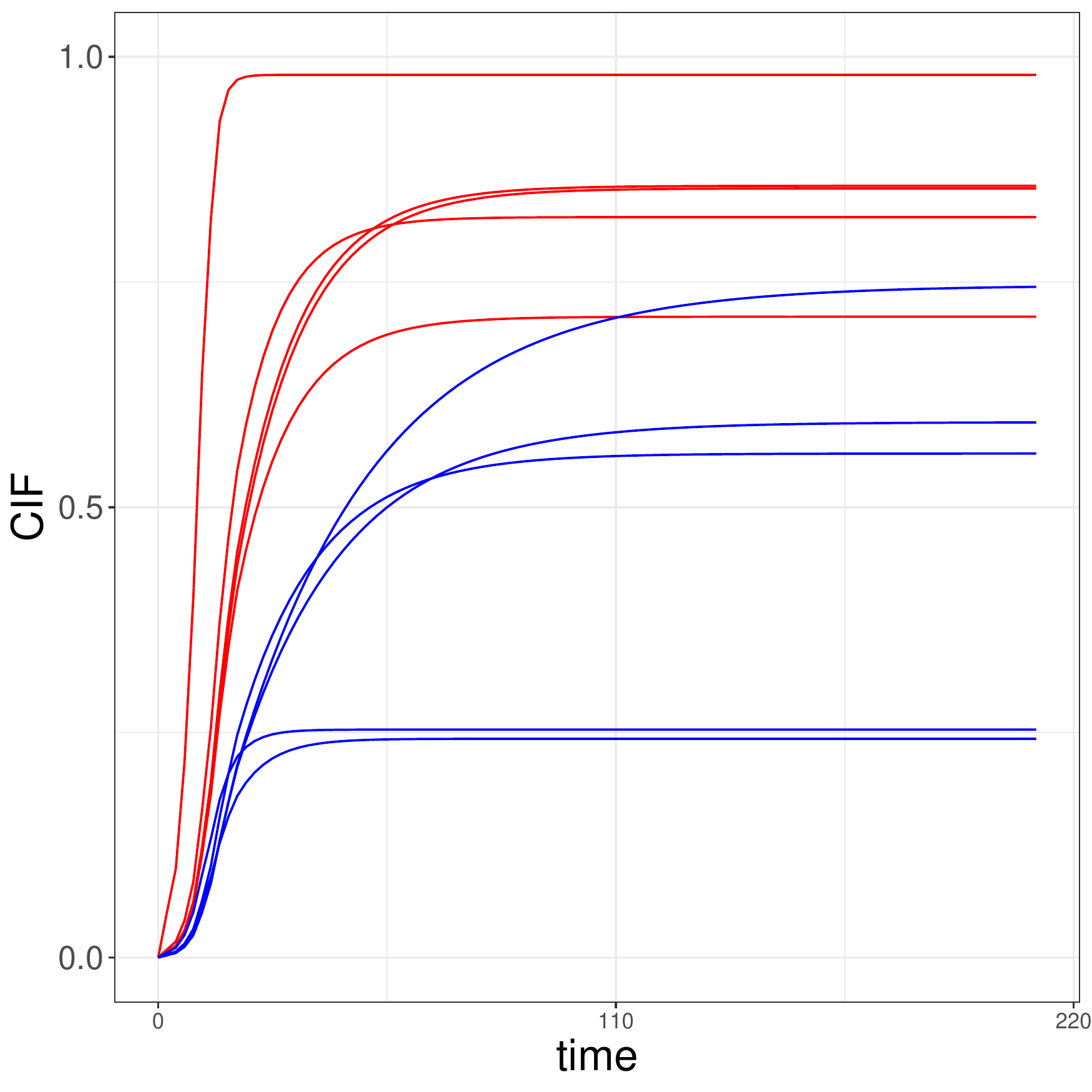} &
\includegraphics[page=3,width=0.45\textwidth]{figure_sub/duo_cif.pdf} \\
\multicolumn{2}{c}{Incorrect Responses}\\
(c) Item 2 & (d) Item 15 \\
\includegraphics[page=2,width=0.45\textwidth]{figure_sub/duo_cif.pdf} &
\includegraphics[page=4,width=0.45\textwidth]{figure_sub/duo_cif.pdf} \\
\end{tabular}
\caption{Cumulative incidence functions (CIFs) of selected respondents and items. We select Item 2 from the blue cluster and Item 15 from the red. Each of the 5 respondents is chosen nearby their cluster centers, and marked by their cluster membership colors. We present correct response CIFs in the top panel and incorrect response CIFs in the bottom.
}
\label{fig:cif_duolingo}
\end{figure}

Figure \ref{fig:cif_duolingo} shows the CIFs for the selected items when the responses are correct (top row) and incorrect (bottom row). The CIFs for the five selected respondents are drawn in each plot, where the line colors indicate their cluster membership. As in the ACT example, we focus on interpreting the CIFs for correct responses below.

In Figure \ref{fig:cif_duolingo}, the blue CIFs tend to show lower peaks than the red CIFs. There are some individual differences among the respondents in reaching the peaks. For example, in the blue group, as the peak is high, the peaks are reached more slowly than when the peak is low. This means that to give a correct response, respondents would spend a long time on solving the items. For item 15, the peaks are generally lower than item 2 for respondents in both groups. The blue CIFs are similar for the two items in terms of peaks and the time to reach the peaks. However, the red CIFs show quite different patterns for the two items. The CIF of one respondent is peculiar and shows a different pattern than other respondents in the same group. Similar to the ACT example, these results suggest that examining CIFs offers valuable and additional insights into the response processes of individual test-takers.

\paragraph*{Model Fit: Posterior Predictive Checks} 

\begin{figure}[htbp]
\centering
\includegraphics[page=1,width=0.32\columnwidth]{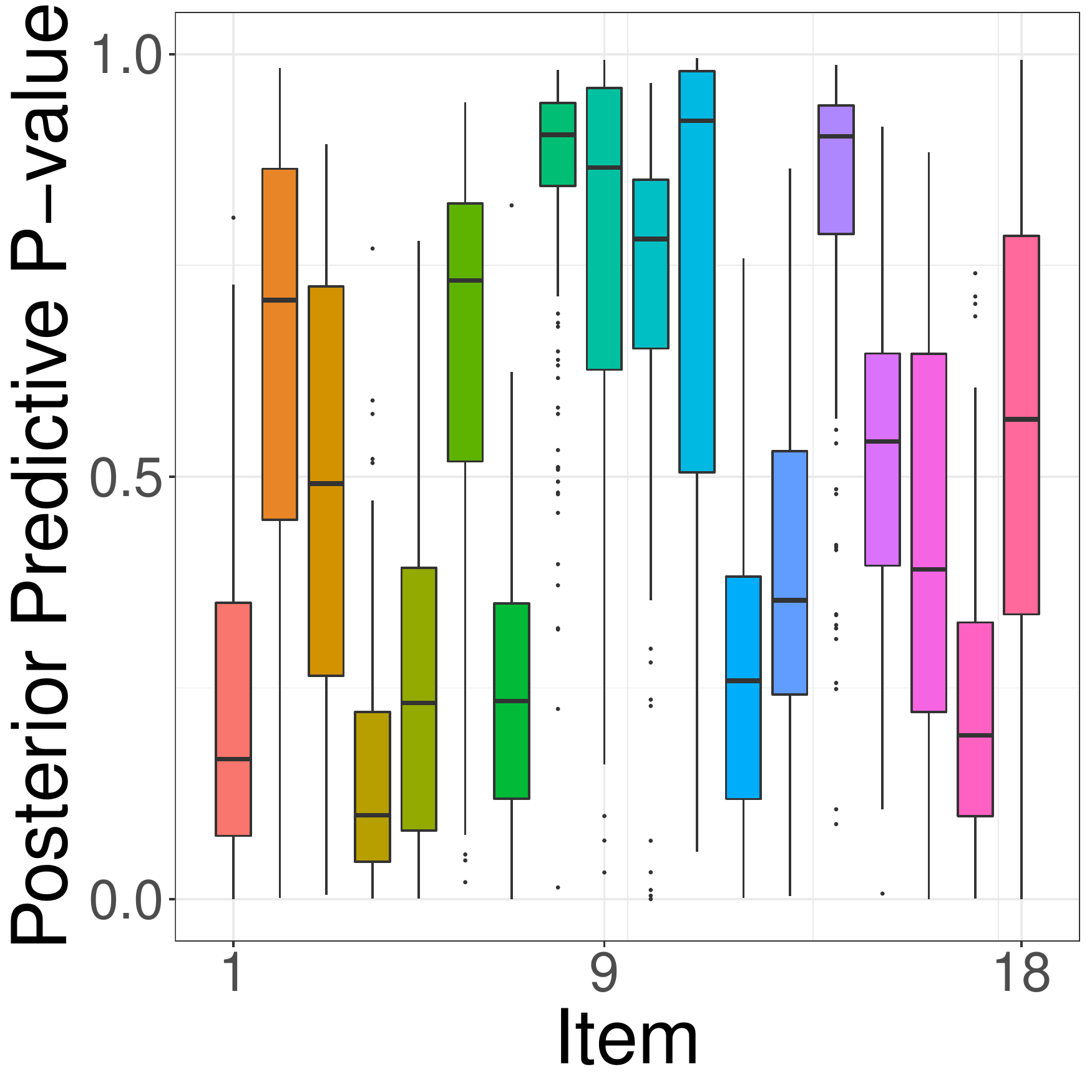}
\includegraphics[page=1,width=0.32\columnwidth]{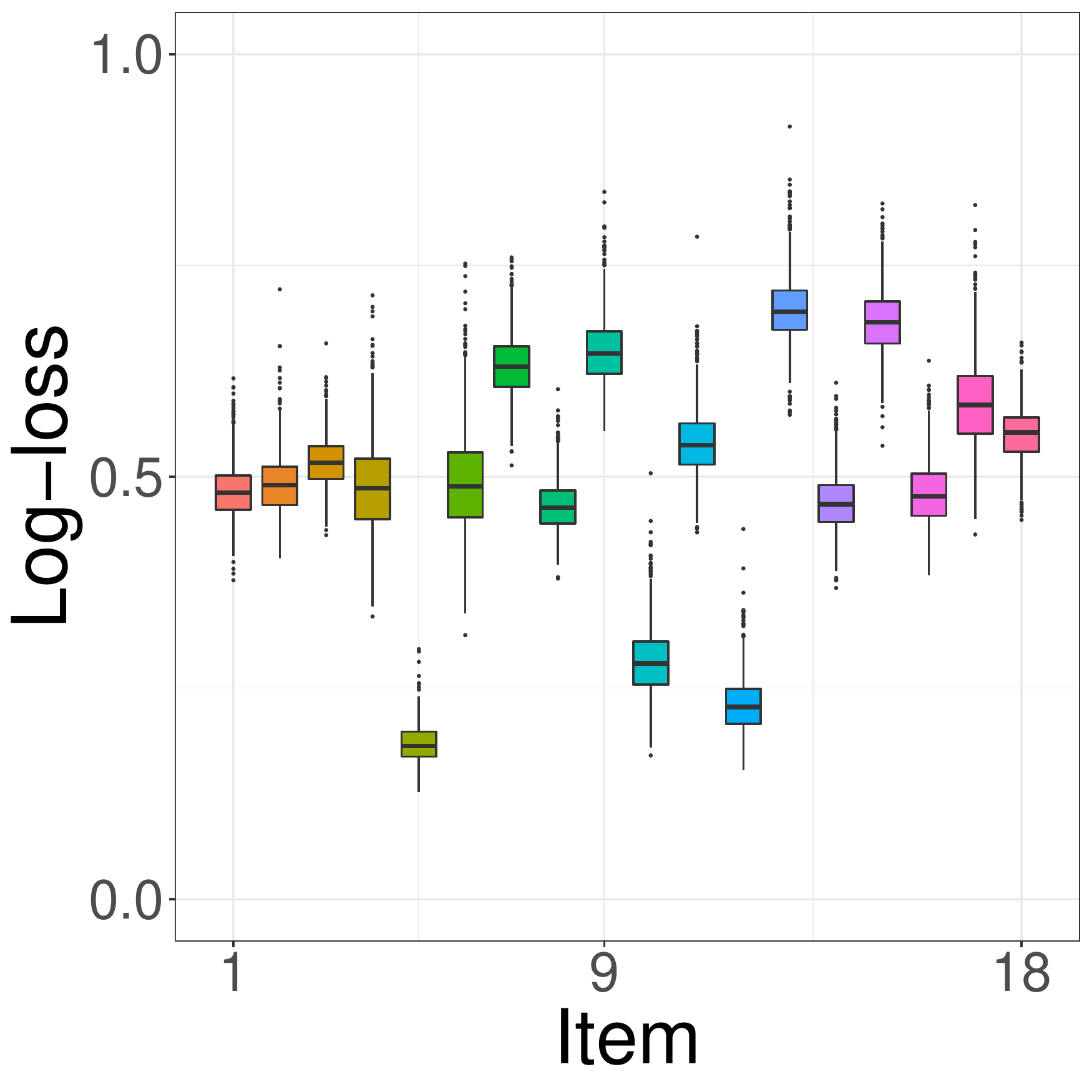}
\includegraphics[page=2,width=0.32\columnwidth]{figure_sub/duolingo_pp_type.pdf}
\caption{\label{fig:duo_pp}
Model fit assessment for Duolingo data. Posterior predictive samples are generated to calculate posterior predictive p-values (left panel), the log-loss (middle panel), and AUC (right panel) for each item.
}
\end{figure}

We evaluate the model fit for the Duolingo data based on \(1,000\) posterior predictive samples. Figure \ref{fig:duo_pp} shows posterior predictive p-values \eqref{eq:ppp}, log-loss values \eqref{eq:logloss}, and AUCs for the data. The posterior predictive p-values appear less desirable than the ACT data example. This may be due to the large variation in the response times, which stems from the no-time limit given in this app-based assessment. We also assume response times above 200 seconds are censored. Thus, simulating response times close to the original data may be more challenging than in the ACT case. The log-loss values and AUCs range between 0.25 and 0.75 and 0.5 and 0.75, respectively, and we concluded that the fit of the proposed model is acceptable for this dataset.

\section{Conclusions}
\label{sec:conclusions}

In this paper, we presented the latent space accumulator model, a new version of the proportional hazard model for cognitive assessment data based on two accumulators corresponding to two competing response outcomes (correct vs. incorrect responses). The proposed model expands \cite{Ranger:2014}'s accumulator model by allowing for dependence between respondents and items in the hazard function per accumulator in the form of distances between respondents and items in a two-dimensional latent space, called an interaction map. With two real data examples, we illustrated how the information from the proposed model, such as the accumulation rates and the cumulative incidence functions, could help improve our understanding of the differences in the item solution processes between correct and incorrect responses. Importantly, we showed how the estimated interaction maps could provide additional insights into the heterogeneity in the accumulation rates across respondents and items.

The proposed framework can be extended in many ways for further generalization. For example, one interesting direction is to extend the proposed approach to settings where more than two response outcomes are available. Such an approach can be beneficial in assessment settings where multiple-choice items involve meaningful distractors. Often selecting different distractors rather than the correct option indicates different levels or types of cognitive functions. In this case, the extended framework will enable us to investigate how different distractors can shape the dependence structures in response times between respondents and test items. This can show how individual test-takers interact differently with test items depending on their cognitive function types and levels. Such analysis can meaningfully improve our understanding of individual differences in the item solution process of cognitive functions. 

\section*{Acknowledgements}

We thank the editor, associate editor, and reviewers for their constructive comments. This study was partially supported by the Yonsei University Research Fund 2022-22-043 and by the Basic Science Research Program through the National Research Foundation of Korea (NRF 2020R1A2C1A01009881). Jin and Yun are co-first authors. 

\bibliography{reference, tempest}

\begin{thebibliography}{}

\bibitem [\protect \citeauthoryear {%
Andersen%
, Geskus%
, Witte%
\BCBL {}\ \BBA {} Putter%
}{%
Andersen%
\ \protect \BOthers {.}}{%
{\protect \APACyear {2012}}%
}]{%
Andersen:2012}
\APACinsertmetastar {%
Andersen:2012}%
\begin{APACrefauthors}%
Andersen, P.%
, Geskus, R.%
, Witte, T.%
\BCBL {}\ \BBA {} Putter, H.%
\end{APACrefauthors}%
\unskip\
\newblock
\APACrefYearMonthDay{2012}{}{}.
\newblock
{\BBOQ}\APACrefatitle {Competing risks in epidemiology: possibilities and
  pitfalls} {Competing risks in epidemiology: possibilities and
  pitfalls}.{\BBCQ}
\newblock
\APACjournalVolNumPages{International Journal of Epidemiology}{41}{}{861-870}.
\PrintBackRefs{\CurrentBib}

\bibitem [\protect \citeauthoryear {%
Austin%
, Lee%
\BCBL {}\ \BBA {} Fine%
}{%
Austin%
\ \protect \BOthers {.}}{%
{\protect \APACyear {2016}}%
}]{%
Austin:2016}
\APACinsertmetastar {%
Austin:2016}%
\begin{APACrefauthors}%
Austin, P\BPBI C.%
, Lee, D\BPBI S.%
\BCBL {}\ \BBA {} Fine, J\BPBI P.%
\end{APACrefauthors}%
\unskip\
\newblock
\APACrefYearMonthDay{2016}{}{}.
\newblock
{\BBOQ}\APACrefatitle {Introduction to the Analysis of Survival Data in the
  Presence of Competing Risks} {Introduction to the analysis of survival data
  in the presence of competing risks}.{\BBCQ}
\newblock
\APACjournalVolNumPages{Circulation}{133}{}{601–609}.
\PrintBackRefs{\CurrentBib}

\bibitem [\protect \citeauthoryear {%
Brown%
\ \BBA {} Heathcote%
}{%
Brown%
\ \BBA {} Heathcote%
}{%
{\protect \APACyear {2005}}%
}]{%
Brown:2005}
\APACinsertmetastar {%
Brown:2005}%
\begin{APACrefauthors}%
Brown, S\BPBI D.%
\BCBT {}\ \BBA {} Heathcote, A.%
\end{APACrefauthors}%
\unskip\
\newblock
\APACrefYearMonthDay{2005}{}{}.
\newblock
{\BBOQ}\APACrefatitle {A ballistic model for choice response times} {A
  ballistic model for choice response times}.{\BBCQ}
\newblock
\APACjournalVolNumPages{Psychological Review}{112}{}{117–128}.
\PrintBackRefs{\CurrentBib}

\bibitem [\protect \citeauthoryear {%
Brown%
\ \BBA {} Heathcote%
}{%
Brown%
\ \BBA {} Heathcote%
}{%
{\protect \APACyear {2008}}%
}]{%
Brown:2008}
\APACinsertmetastar {%
Brown:2008}%
\begin{APACrefauthors}%
Brown, S\BPBI D.%
\BCBT {}\ \BBA {} Heathcote, A.%
\end{APACrefauthors}%
\unskip\
\newblock
\APACrefYearMonthDay{2008}{}{}.
\newblock
{\BBOQ}\APACrefatitle {The simplest complete model of choice response time:
  Linear ballistic accumulation} {The simplest complete model of choice
  response time: Linear ballistic accumulation}.{\BBCQ}
\newblock
\APACjournalVolNumPages{Cognitive Psychology}{57}{}{153–178}.
\PrintBackRefs{\CurrentBib}

\bibitem [\protect \citeauthoryear {%
Carpenter%
\ \protect \BOthers {.}}{%
Carpenter%
\ \protect \BOthers {.}}{%
{\protect \APACyear {2017}}%
}]{%
Carpenter:17}
\APACinsertmetastar {%
Carpenter:17}%
\begin{APACrefauthors}%
Carpenter, B.%
, Gelman, A.%
, Hoffman, M\BPBI D.%
, Lee, D.%
, Goodrich, B.%
, Betancourt, M.%
\BDBL {}Riddell, A.%
\end{APACrefauthors}%
\unskip\
\newblock
\APACrefYearMonthDay{2017}{}{}.
\newblock
{\BBOQ}\APACrefatitle {Stan: {A} probabilistic programming language} {Stan: {A}
  probabilistic programming language}.{\BBCQ}
\newblock
\APACjournalVolNumPages{Journal of Statistical Software}{76}{}{}.
\PrintBackRefs{\CurrentBib}

\bibitem [\protect \citeauthoryear {%
De~Boeck%
\ \BBA {} Jeon%
}{%
De~Boeck%
\ \BBA {} Jeon%
}{%
{\protect \APACyear {2019}}%
}]{%
DeBoeck:2019}
\APACinsertmetastar {%
DeBoeck:2019}%
\begin{APACrefauthors}%
De~Boeck, P.%
\BCBT {}\ \BBA {} Jeon, M.%
\end{APACrefauthors}%
\unskip\
\newblock
\APACrefYearMonthDay{2019}{}{}.
\newblock
{\BBOQ}\APACrefatitle {An Overview of Models for Response Times and Processes
  in Cognitive Tests} {An overview of models for response times and processes
  in cognitive tests}.{\BBCQ}
\newblock
\APACjournalVolNumPages{Frontiers in psychology}{10}{}{}.
\PrintBackRefs{\CurrentBib}

\bibitem [\protect \citeauthoryear {%
Dhillon%
}{%
Dhillon%
}{%
{\protect \APACyear {2001}}%
}]{%
Dhillon:2001}
\APACinsertmetastar {%
Dhillon:2001}%
\begin{APACrefauthors}%
Dhillon, I\BPBI S.%
\end{APACrefauthors}%
\unskip\
\newblock
\APACrefYearMonthDay{2001}{}{}.
\newblock
{\BBOQ}\APACrefatitle {Co-clustering documents and words using bipartite
  spectral graph partitioning} {Co-clustering documents and words using
  bipartite spectral graph partitioning}.{\BBCQ}
\newblock
\BIn{} \APACrefbtitle {Proceedings of the seventh ACM SIGKDD international
  conference on Knowledge discovery and data mining} {Proceedings of the
  seventh acm sigkdd international conference on knowledge discovery and data
  mining}\ (\BPGS\ 269--274).
\newblock
\APACaddressPublisher{New York, NY, USA}{}.
\PrintBackRefs{\CurrentBib}

\bibitem [\protect \citeauthoryear {%
Douglas%
, Kosorok%
\BCBL {}\ \BBA {} Chewing%
}{%
Douglas%
\ \protect \BOthers {.}}{%
{\protect \APACyear {1999}}%
}]{%
Douglas:1999}
\APACinsertmetastar {%
Douglas:1999}%
\begin{APACrefauthors}%
Douglas, J.%
, Kosorok, M.%
\BCBL {}\ \BBA {} Chewing, B.%
\end{APACrefauthors}%
\unskip\
\newblock
\APACrefYearMonthDay{1999}{}{}.
\newblock
{\BBOQ}\APACrefatitle {A latent variable model for discrete multivariate
  psychometric waiting times} {A latent variable model for discrete
  multivariate psychometric waiting times}.{\BBCQ}
\newblock
\APACjournalVolNumPages{Psychometrika}{64}{}{69-82}.
\PrintBackRefs{\CurrentBib}

\bibitem [\protect \citeauthoryear {%
Fazio%
}{%
Fazio%
}{%
{\protect \APACyear {1995}}%
}]{%
Fazio:1995}
\APACinsertmetastar {%
Fazio:1995}%
\begin{APACrefauthors}%
Fazio, R.%
\end{APACrefauthors}%
\unskip\
\newblock
\APACrefYearMonthDay{1995}{}{}.
\newblock
{\BBOQ}\APACrefatitle {Attitudes as object-evaluation associations:
  Determinants, consequences, and correlates of attitude accessibility}
  {Attitudes as object-evaluation associations: Determinants, consequences, and
  correlates of attitude accessibility}.{\BBCQ}
\newblock
\BIn{} R.~Petty\ \BBA {} J.~Krosnick\ (\BEDS), \APACrefbtitle {Attitude
  strength: Antecedents and consequences} {Attitude strength: Antecedents and
  consequences}\ (\BPG~247–282).
\newblock
\APACaddressPublisher{Mahwah, NJ}{Erlbaum}.
\PrintBackRefs{\CurrentBib}

\bibitem [\protect \citeauthoryear {%
Fine%
\ \BBA {} Gray%
}{%
Fine%
\ \BBA {} Gray%
}{%
{\protect \APACyear {1999}}%
}]{%
Fine:1999}
\APACinsertmetastar {%
Fine:1999}%
\begin{APACrefauthors}%
Fine, J.%
\BCBT {}\ \BBA {} Gray, R.%
\end{APACrefauthors}%
\unskip\
\newblock
\APACrefYearMonthDay{1999}{}{}.
\newblock
{\BBOQ}\APACrefatitle {A proportional hazards model for the subdistribution of
  a competing risk} {A proportional hazards model for the subdistribution of a
  competing risk}.{\BBCQ}
\newblock
\APACjournalVolNumPages{Journal of the American Statistical
  Association}{94}{}{496-509}.
\PrintBackRefs{\CurrentBib}

\bibitem [\protect \citeauthoryear {%
Friel%
, Rastelli%
, Wyse%
\BCBL {}\ \BBA {} Raftery%
}{%
Friel%
\ \protect \BOthers {.}}{%
{\protect \APACyear {2016}}%
}]{%
Friel:2016}
\APACinsertmetastar {%
Friel:2016}%
\begin{APACrefauthors}%
Friel, N.%
, Rastelli, R.%
, Wyse, J.%
\BCBL {}\ \BBA {} Raftery, A\BPBI E.%
\end{APACrefauthors}%
\unskip\
\newblock
\APACrefYearMonthDay{2016}{}{}.
\newblock
{\BBOQ}\APACrefatitle {Interlocking directorates in {I}rish companies using a
  latent space model for bipartite networks} {Interlocking directorates in
  {I}rish companies using a latent space model for bipartite networks}.{\BBCQ}
\newblock
\APACjournalVolNumPages{Proceedings of the National Academy of Sciences of the
  United States of America}{113}{}{6629-6634}.
\PrintBackRefs{\CurrentBib}

\bibitem [\protect \citeauthoryear {%
Gelman%
\ \BBA {} Rubin%
}{%
Gelman%
\ \BBA {} Rubin%
}{%
{\protect \APACyear {1992}}%
}]{%
Gelman:1992}
\APACinsertmetastar {%
Gelman:1992}%
\begin{APACrefauthors}%
Gelman, A.%
\BCBT {}\ \BBA {} Rubin, D\BPBI B.%
\end{APACrefauthors}%
\unskip\
\newblock
\APACrefYearMonthDay{1992}{}{}.
\newblock
{\BBOQ}\APACrefatitle {Inference from iterative simulation using multiple
  sequences.} {Inference from iterative simulation using multiple
  sequences.}{\BBCQ}
\newblock
\APACjournalVolNumPages{Statistical Science}{7}{}{457-472}.
\PrintBackRefs{\CurrentBib}

\bibitem [\protect \citeauthoryear {%
Goldhammer%
}{%
Goldhammer%
}{%
{\protect \APACyear {2015}}%
}]{%
Goldhammer:2015}
\APACinsertmetastar {%
Goldhammer:2015}%
\begin{APACrefauthors}%
Goldhammer, F.%
\end{APACrefauthors}%
\unskip\
\newblock
\APACrefYearMonthDay{2015}{}{}.
\newblock
{\BBOQ}\APACrefatitle {Measuring Ability, Speed, or Both? Challenges,
  Psychometric Solutions, and What Can Be Gained From Experimental Control}
  {Measuring ability, speed, or both? challenges, psychometric solutions, and
  what can be gained from experimental control}.{\BBCQ}
\newblock
\APACjournalVolNumPages{Measurement: interdisciplinary research and
  perspectives}{13(3-4)}{}{133-164}.
\PrintBackRefs{\CurrentBib}

\bibitem [\protect \citeauthoryear {%
Gower%
}{%
Gower%
}{%
{\protect \APACyear {1975}}%
}]{%
Gower:1975}
\APACinsertmetastar {%
Gower:1975}%
\begin{APACrefauthors}%
Gower, J\BPBI C.%
\end{APACrefauthors}%
\unskip\
\newblock
\APACrefYearMonthDay{1975}{}{}.
\newblock
{\BBOQ}\APACrefatitle {Generalized procrustes analysis} {Generalized procrustes
  analysis}.{\BBCQ}
\newblock
\APACjournalVolNumPages{Psychometrika}{40}{}{33-51}.
\PrintBackRefs{\CurrentBib}

\bibitem [\protect \citeauthoryear {%
Handcock%
, Raftery%
\BCBL {}\ \BBA {} Tantrum%
}{%
Handcock%
\ \protect \BOthers {.}}{%
{\protect \APACyear {2007}}%
}]{%
Handcock:2007}
\APACinsertmetastar {%
Handcock:2007}%
\begin{APACrefauthors}%
Handcock, M\BPBI S.%
, Raftery, A\BPBI E.%
\BCBL {}\ \BBA {} Tantrum, J\BPBI M.%
\end{APACrefauthors}%
\unskip\
\newblock
\APACrefYearMonthDay{2007}{}{}.
\newblock
{\BBOQ}\APACrefatitle {Model-based clustering for social network} {Model-based
  clustering for social network}.{\BBCQ}
\newblock
\APACjournalVolNumPages{Journal of the Royal Statistical Society, Series
  A}{170}{}{301-354}.
\PrintBackRefs{\CurrentBib}

\bibitem [\protect \citeauthoryear {%
Hoff%
, Raftery%
\BCBL {}\ \BBA {} Handcock%
}{%
Hoff%
\ \protect \BOthers {.}}{%
{\protect \APACyear {2002}}%
}]{%
Hoff:2002}
\APACinsertmetastar {%
Hoff:2002}%
\begin{APACrefauthors}%
Hoff, P.%
, Raftery, A.%
\BCBL {}\ \BBA {} Handcock, M\BPBI S.%
\end{APACrefauthors}%
\unskip\
\newblock
\APACrefYearMonthDay{2002}{}{}.
\newblock
{\BBOQ}\APACrefatitle {Latent space approaches to social network analysis}
  {Latent space approaches to social network analysis}.{\BBCQ}
\newblock
\APACjournalVolNumPages{Journal of the American Statistical
  Association}{97}{}{1090-1098}.
\PrintBackRefs{\CurrentBib}

\bibitem [\protect \citeauthoryear {%
Ibrahim%
, Chen%
\BCBL {}\ \BBA {} Sinha%
}{%
Ibrahim%
\ \protect \BOthers {.}}{%
{\protect \APACyear {2001}}%
}]{%
Ibrahim:2001}
\APACinsertmetastar {%
Ibrahim:2001}%
\begin{APACrefauthors}%
Ibrahim, J\BPBI G.%
, Chen, M\BHBI H.%
\BCBL {}\ \BBA {} Sinha, D.%
\end{APACrefauthors}%
\unskip\
\newblock
\APACrefYear{2001}.
\newblock
\APACrefbtitle {Bayesian Survival Analysis} {Bayesian survival analysis}.
\newblock
\APACaddressPublisher{New York}{Springer}.
\PrintBackRefs{\CurrentBib}

\bibitem [\protect \citeauthoryear {%
Jansen%
}{%
Jansen%
}{%
{\protect \APACyear {1997}}%
}]{%
Jansen:1997}
\APACinsertmetastar {%
Jansen:1997}%
\begin{APACrefauthors}%
Jansen, M.%
\end{APACrefauthors}%
\unskip\
\newblock
\APACrefYearMonthDay{1997}{}{}.
\newblock
{\BBOQ}\APACrefatitle {Rasch’s model for reading speed with manifest
  explanatory variables} {Rasch’s model for reading speed with manifest
  explanatory variables}.{\BBCQ}
\newblock
\APACjournalVolNumPages{Psychometrika}{62}{}{393-409}.
\PrintBackRefs{\CurrentBib}

\bibitem [\protect \citeauthoryear {%
Jeon%
, Jin%
, Schweinberger%
\BCBL {}\ \BBA {} Baugh%
}{%
Jeon%
\ \protect \BOthers {.}}{%
{\protect \APACyear {2021}}%
}]{%
Jeon:2021}
\APACinsertmetastar {%
Jeon:2021}%
\begin{APACrefauthors}%
Jeon, M.%
, Jin, I\BPBI H.%
, Schweinberger, M.%
\BCBL {}\ \BBA {} Baugh, S.%
\end{APACrefauthors}%
\unskip\
\newblock
\APACrefYearMonthDay{2021}{}{}.
\newblock
{\BBOQ}\APACrefatitle {Mapping Unobserved Item–Respondent Interactions: A
  Latent Space Item Response Model with Interaction Map} {Mapping unobserved
  item–respondent interactions: A latent space item response model with
  interaction map}.{\BBCQ}
\newblock
\APACjournalVolNumPages{Psychometrika}{86}{}{378-403}.
\PrintBackRefs{\CurrentBib}

\bibitem [\protect \citeauthoryear {%
Jin%
, Liu%
, Thall%
\BCBL {}\ \BBA {} Yuan%
}{%
Jin%
\ \protect \BOthers {.}}{%
{\protect \APACyear {2014}}%
}]{%
Jin:2014JASA}
\APACinsertmetastar {%
Jin:2014JASA}%
\begin{APACrefauthors}%
Jin, I.%
, Liu, S.%
, Thall, P.%
\BCBL {}\ \BBA {} Yuan, Y.%
\end{APACrefauthors}%
\unskip\
\newblock
\APACrefYearMonthDay{2014}{}{}.
\newblock
{\BBOQ}\APACrefatitle {Using data augmentation to facilitate conduct of phase
  {I-II} clinical trials with delayed outcome} {Using data augmentation to
  facilitate conduct of phase {I-II} clinical trials with delayed
  outcome}.{\BBCQ}
\newblock
\APACjournalVolNumPages{Journal of the American Statistical
  Association}{109}{}{525-536}.
\PrintBackRefs{\CurrentBib}

\bibitem [\protect \citeauthoryear {%
Kang%
}{%
Kang%
}{%
{\protect \APACyear {2016}}%
}]{%
Kang:2016}
\APACinsertmetastar {%
Kang:2016}%
\begin{APACrefauthors}%
Kang, H.%
\end{APACrefauthors}%
\unskip\
\newblock
\APACrefYearMonthDay{2016}{}{}.
\newblock
{\BBOQ}\APACrefatitle {Penalized partial likelihood inference of proportional
  hazards latent trait models} {Penalized partial likelihood inference of
  proportional hazards latent trait models}.{\BBCQ}
\newblock
\APACjournalVolNumPages{British Journal of Mathematical and Statistical
  Psychology}{70(2)}{}{187-208}.
\PrintBackRefs{\CurrentBib}

\bibitem [\protect \citeauthoryear {%
Krivitsky%
, Handcock%
, Raftery%
\BCBL {}\ \BBA {} Hoff%
}{%
Krivitsky%
\ \protect \BOthers {.}}{%
{\protect \APACyear {2009}}%
}]{%
Krivitsky:2009}
\APACinsertmetastar {%
Krivitsky:2009}%
\begin{APACrefauthors}%
Krivitsky, P\BPBI N.%
, Handcock, M\BPBI S.%
, Raftery, A\BPBI E.%
\BCBL {}\ \BBA {} Hoff, P\BPBI D.%
\end{APACrefauthors}%
\unskip\
\newblock
\APACrefYearMonthDay{2009}{}{}.
\newblock
{\BBOQ}\APACrefatitle {Representing degree distributions, clustering, and
  homophily in social networks with latent cluster random network models}
  {Representing degree distributions, clustering, and homophily in social
  networks with latent cluster random network models}.{\BBCQ}
\newblock
\APACjournalVolNumPages{Social Networks}{31}{}{204-213}.
\PrintBackRefs{\CurrentBib}

\bibitem [\protect \citeauthoryear {%
Kyllonen%
\ \BBA {} Zu%
}{%
Kyllonen%
\ \BBA {} Zu%
}{%
{\protect \APACyear {2016}}%
}]{%
Kyllonen:2016}
\APACinsertmetastar {%
Kyllonen:2016}%
\begin{APACrefauthors}%
Kyllonen, P.%
\BCBT {}\ \BBA {} Zu, J.%
\end{APACrefauthors}%
\unskip\
\newblock
\APACrefYearMonthDay{2016}{}{}.
\newblock
{\BBOQ}\APACrefatitle {Use of Response Time for Measuring Cognitive Ability}
  {Use of response time for measuring cognitive ability}.{\BBCQ}
\newblock
\APACjournalVolNumPages{Journal of Intelligence}{4(4)}{}{14}.
\PrintBackRefs{\CurrentBib}

\bibitem [\protect \citeauthoryear {%
Lau%
, Cole%
\BCBL {}\ \BBA {} Gange%
}{%
Lau%
\ \protect \BOthers {.}}{%
{\protect \APACyear {2009}}%
}]{%
Lau:2009}
\APACinsertmetastar {%
Lau:2009}%
\begin{APACrefauthors}%
Lau, B.%
, Cole, S.%
\BCBL {}\ \BBA {} Gange, S.%
\end{APACrefauthors}%
\unskip\
\newblock
\APACrefYearMonthDay{2009}{}{}.
\newblock
{\BBOQ}\APACrefatitle {Competing risk regression models for epidemiologic data}
  {Competing risk regression models for epidemiologic data}.{\BBCQ}
\newblock
\APACjournalVolNumPages{American journal of epidemiology}{170}{}{244-256}.
\PrintBackRefs{\CurrentBib}

\bibitem [\protect \citeauthoryear {%
Lee%
\ \BBA {} Chen%
}{%
Lee%
\ \BBA {} Chen%
}{%
{\protect \APACyear {2011}}%
}]{%
Lee:2011}
\APACinsertmetastar {%
Lee:2011}%
\begin{APACrefauthors}%
Lee, Y\BPBI H.%
\BCBT {}\ \BBA {} Chen, H.%
\end{APACrefauthors}%
\unskip\
\newblock
\APACrefYearMonthDay{2011}{}{}.
\newblock
{\BBOQ}\APACrefatitle {A review of recent response-time analyses in educational
  testing} {A review of recent response-time analyses in educational
  testing}.{\BBCQ}
\newblock
\APACjournalVolNumPages{Psychological Test and Assessment
  Modeling}{53(3)}{}{359}.
\PrintBackRefs{\CurrentBib}

\bibitem [\protect \citeauthoryear {%
Liu%
\ \BBA {} Wang%
}{%
Liu%
\ \BBA {} Wang%
}{%
{\protect \APACyear {2022}}%
}]{%
Liu:2022}
\APACinsertmetastar {%
Liu:2022}%
\begin{APACrefauthors}%
Liu, Y.%
\BCBT {}\ \BBA {} Wang, W.%
\end{APACrefauthors}%
\unskip\
\newblock
\APACrefYearMonthDay{2022}{}{}.
\newblock
{\BBOQ}\APACrefatitle {Semiparametric Factor Analysis for Item-Level Response
  Time Data} {Semiparametric factor analysis for item-level response time
  data}.{\BBCQ}
\newblock
\APACjournalVolNumPages{Psychometrika}{87(2)}{}{666-692}.
\PrintBackRefs{\CurrentBib}

\bibitem [\protect \citeauthoryear {%
Loeys%
, Legrand%
, Schettino%
\BCBL {}\ \BBA {} Pourtois%
}{%
Loeys%
\ \protect \BOthers {.}}{%
{\protect \APACyear {2014}}%
}]{%
Loeys:2014}
\APACinsertmetastar {%
Loeys:2014}%
\begin{APACrefauthors}%
Loeys, T.%
, Legrand, C.%
, Schettino, A.%
\BCBL {}\ \BBA {} Pourtois, G.%
\end{APACrefauthors}%
\unskip\
\newblock
\APACrefYearMonthDay{2014}{}{}.
\newblock
{\BBOQ}\APACrefatitle {Semi-parametric proportional hazards models with crossed
  random effects for psychometric response times} {Semi-parametric proportional
  hazards models with crossed random effects for psychometric response
  times}.{\BBCQ}
\newblock
\APACjournalVolNumPages{British Journal of Mathematical and Statistical
  Psychology}{67(2)}{}{304-327}.
\PrintBackRefs{\CurrentBib}

\bibitem [\protect \citeauthoryear {%
Maris%
}{%
Maris%
}{%
{\protect \APACyear {1993}}%
}]{%
Maris:1993}
\APACinsertmetastar {%
Maris:1993}%
\begin{APACrefauthors}%
Maris, E.%
\end{APACrefauthors}%
\unskip\
\newblock
\APACrefYearMonthDay{1993}{}{}.
\newblock
{\BBOQ}\APACrefatitle {Additive and multiplicative models for gamma distributed
  random variables, and their application as psychometric models for response
  times} {Additive and multiplicative models for gamma distributed random
  variables, and their application as psychometric models for response
  times}.{\BBCQ}
\newblock
\APACjournalVolNumPages{Psychometrika}{58}{}{445-469}.
\PrintBackRefs{\CurrentBib}

\bibitem [\protect \citeauthoryear {%
Molenaar%
, Bolsinova%
\BCBL {}\ \BBA {} Vermunt%
}{%
Molenaar%
\ \protect \BOthers {.}}{%
{\protect \APACyear {2018}}%
}]{%
Molenaar:2018}
\APACinsertmetastar {%
Molenaar:2018}%
\begin{APACrefauthors}%
Molenaar, D.%
, Bolsinova, M.%
\BCBL {}\ \BBA {} Vermunt, J.%
\end{APACrefauthors}%
\unskip\
\newblock
\APACrefYearMonthDay{2018}{}{}.
\newblock
{\BBOQ}\APACrefatitle {A semi-parametric within-subject mixture approach to the
  analyses of responses and response times} {A semi-parametric within-subject
  mixture approach to the analyses of responses and response times}.{\BBCQ}
\newblock
\APACjournalVolNumPages{British Journal of Mathematical and Statistical
  Psychology}{71(2)}{}{205-228}.
\PrintBackRefs{\CurrentBib}

\bibitem [\protect \citeauthoryear {%
{{R Core Team}}%
}{%
{{R Core Team}}%
}{%
{\protect \APACyear {{2020}}}%
}]{%
r_core_team_r_2020}
\APACinsertmetastar {%
r_core_team_r_2020}%
\begin{APACrefauthors}%
{{R Core Team}}.%
\end{APACrefauthors}%
\unskip\
\newblock
\APACrefYearMonthDay{{2020}}{}{}.
\newblock
\APACrefbtitle {{R: A Language and Environment for Statistical Computing}} {{R:
  A Language and Environment for Statistical Computing}}\ [{Manual}].
\newblock
\APACaddressPublisher{{{Vienna, Austria}}}{}.
\PrintBackRefs{\CurrentBib}

\bibitem [\protect \citeauthoryear {%
Raftery%
, Niu%
, Hoff%
\BCBL {}\ \BBA {} Yeung%
}{%
Raftery%
\ \protect \BOthers {.}}{%
{\protect \APACyear {2012}}%
}]{%
Raftery:2012}
\APACinsertmetastar {%
Raftery:2012}%
\begin{APACrefauthors}%
Raftery, A.%
, Niu, X.%
, Hoff, P.%
\BCBL {}\ \BBA {} Yeung, K.%
\end{APACrefauthors}%
\unskip\
\newblock
\APACrefYearMonthDay{2012}{}{}.
\newblock
{\BBOQ}\APACrefatitle {Fast inference for the latent space network model using
  a case-control approximate likelihood} {Fast inference for the latent space
  network model using a case-control approximate likelihood}.{\BBCQ}
\newblock
\APACjournalVolNumPages{Journal of Computational and Graphical
  Statistics}{21}{}{909-919}.
\PrintBackRefs{\CurrentBib}

\bibitem [\protect \citeauthoryear {%
Ranger%
\ \BBA {} Kuhn%
}{%
Ranger%
\ \BBA {} Kuhn%
}{%
{\protect \APACyear {2012}}%
}]{%
RangerK:2012}
\APACinsertmetastar {%
RangerK:2012}%
\begin{APACrefauthors}%
Ranger, J.%
\BCBT {}\ \BBA {} Kuhn, J.%
\end{APACrefauthors}%
\unskip\
\newblock
\APACrefYearMonthDay{2012}{}{}.
\newblock
{\BBOQ}\APACrefatitle {A flexible latent trait model for response times in
  tests} {A flexible latent trait model for response times in tests}.{\BBCQ}
\newblock
\APACjournalVolNumPages{Psychometrika}{77}{}{31-47}.
\PrintBackRefs{\CurrentBib}

\bibitem [\protect \citeauthoryear {%
Ranger%
\ \BBA {} Kuhn%
}{%
Ranger%
\ \BBA {} Kuhn%
}{%
{\protect \APACyear {2014}}%
}]{%
Ranger:2014}
\APACinsertmetastar {%
Ranger:2014}%
\begin{APACrefauthors}%
Ranger, J.%
\BCBT {}\ \BBA {} Kuhn, J.%
\end{APACrefauthors}%
\unskip\
\newblock
\APACrefYearMonthDay{2014}{}{}.
\newblock
{\BBOQ}\APACrefatitle {An Accumulator model for responses and response time in
  tests based on the proportional hazards model} {An accumulator model for
  responses and response time in tests based on the proportional hazards
  model}.{\BBCQ}
\newblock
\APACjournalVolNumPages{British Journal of Mathematical and Statistical
  Psychology}{67}{}{388-407}.
\PrintBackRefs{\CurrentBib}

\bibitem [\protect \citeauthoryear {%
Ranger%
\ \BBA {} Kuhn%
}{%
Ranger%
\ \BBA {} Kuhn%
}{%
{\protect \APACyear {2021}}%
}]{%
Ranger:2021}
\APACinsertmetastar {%
Ranger:2021}%
\begin{APACrefauthors}%
Ranger, J.%
\BCBT {}\ \BBA {} Kuhn, J.%
\end{APACrefauthors}%
\unskip\
\newblock
\APACrefYearMonthDay{2021}{}{}.
\newblock
{\BBOQ}\APACrefatitle {A Semiparametric Latent Trait Model for response Times
  in Tests} {A semiparametric latent trait model for response times in
  tests}.{\BBCQ}
\newblock
\APACjournalVolNumPages{Psychological Test and Assessment
  Modeling}{63(3)}{}{396-431}.
\PrintBackRefs{\CurrentBib}

\bibitem [\protect \citeauthoryear {%
Ranger%
\ \BBA {} Ortner%
}{%
Ranger%
\ \BBA {} Ortner%
}{%
{\protect \APACyear {2012}}%
}]{%
Ranger:2012}
\APACinsertmetastar {%
Ranger:2012}%
\begin{APACrefauthors}%
Ranger, J.%
\BCBT {}\ \BBA {} Ortner, T.%
\end{APACrefauthors}%
\unskip\
\newblock
\APACrefYearMonthDay{2012}{}{}.
\newblock
{\BBOQ}\APACrefatitle {A latent trait model for response times on tests
  employing the proportional hazards model} {A latent trait model for response
  times on tests employing the proportional hazards model}.{\BBCQ}
\newblock
\APACjournalVolNumPages{British Journal of Mathematical and Statistical
  Psychology}{65}{}{334-349}.
\PrintBackRefs{\CurrentBib}

\bibitem [\protect \citeauthoryear {%
Ranger%
\ \BBA {} Ortner%
}{%
Ranger%
\ \BBA {} Ortner%
}{%
{\protect \APACyear {2013}}%
}]{%
Ranger:2013}
\APACinsertmetastar {%
Ranger:2013}%
\begin{APACrefauthors}%
Ranger, J.%
\BCBT {}\ \BBA {} Ortner, T.%
\end{APACrefauthors}%
\unskip\
\newblock
\APACrefYearMonthDay{2013}{}{}.
\newblock
{\BBOQ}\APACrefatitle {Response time modeling based on the proportional hazards
  model} {Response time modeling based on the proportional hazards
  model}.{\BBCQ}
\newblock
\APACjournalVolNumPages{Multivariate Behavioral Research}{48}{}{503-533}.
\PrintBackRefs{\CurrentBib}

\bibitem [\protect \citeauthoryear {%
Rouder%
, Lu%
, Speckman%
, Sun%
\BCBL {}\ \BBA {} Jiang%
}{%
Rouder%
\ \protect \BOthers {.}}{%
{\protect \APACyear {2005}}%
}]{%
Rouder:2005}
\APACinsertmetastar {%
Rouder:2005}%
\begin{APACrefauthors}%
Rouder, J.%
, Lu, J.%
, Speckman, P.%
, Sun, D.%
\BCBL {}\ \BBA {} Jiang, Y.%
\end{APACrefauthors}%
\unskip\
\newblock
\APACrefYearMonthDay{2005}{}{}.
\newblock
{\BBOQ}\APACrefatitle {A hierarchical model for estimating response time
  distributions} {A hierarchical model for estimating response time
  distributions}.{\BBCQ}
\newblock
\APACjournalVolNumPages{Psychometrika}{12(2)}{}{195-223}.
\PrintBackRefs{\CurrentBib}

\bibitem [\protect \citeauthoryear {%
Usher%
\ \BBA {} McClelland%
}{%
Usher%
\ \BBA {} McClelland%
}{%
{\protect \APACyear {2001}}%
}]{%
Usher:2001}
\APACinsertmetastar {%
Usher:2001}%
\begin{APACrefauthors}%
Usher, M.%
\BCBT {}\ \BBA {} McClelland, J\BPBI L.%
\end{APACrefauthors}%
\unskip\
\newblock
\APACrefYearMonthDay{2001}{}{}.
\newblock
{\BBOQ}\APACrefatitle {The time course of perceptual choice: The leaky,
  competing accumulator model} {The time course of perceptual choice: The
  leaky, competing accumulator model}.{\BBCQ}
\newblock
\APACjournalVolNumPages{Psychological Review}{108}{}{550–592}.
\PrintBackRefs{\CurrentBib}

\bibitem [\protect \citeauthoryear {%
{van Breukelen}%
\ \protect \BOthers {.}}{%
{van Breukelen}%
\ \protect \BOthers {.}}{%
{\protect \APACyear {1995}}%
}]{%
VanBreukelen:1995}
\APACinsertmetastar {%
VanBreukelen:1995}%
\begin{APACrefauthors}%
{van Breukelen}, G.%
, Roskam, E.%
, Eling, P.%
, Jansen, R.%
, Souren, D.%
\BCBL {}\ \BBA {} Ickenroth, J.%
\end{APACrefauthors}%
\unskip\
\newblock
\APACrefYearMonthDay{1995}{}{}.
\newblock
{\BBOQ}\APACrefatitle {A model and diagnostic measures for response time series
  on tests of concentration: Historical background, conceptual framework, and
  some applications} {A model and diagnostic measures for response time series
  on tests of concentration: Historical background, conceptual framework, and
  some applications}.{\BBCQ}
\newblock
\APACjournalVolNumPages{Brain and Cognition}{27}{}{147–179}.
\PrintBackRefs{\CurrentBib}

\bibitem [\protect \citeauthoryear {%
{van der Mass}%
\ \BBA {} Wagenmakers%
}{%
{van der Mass}%
\ \BBA {} Wagenmakers%
}{%
{\protect \APACyear {2005}}%
}]{%
vanderMass:2016}
\APACinsertmetastar {%
vanderMass:2016}%
\begin{APACrefauthors}%
{van der Mass}, H\BPBI L\BPBI J.%
\BCBT {}\ \BBA {} Wagenmakers, E\BHBI J.%
\end{APACrefauthors}%
\unskip\
\newblock
\APACrefYearMonthDay{2005}{}{}.
\newblock
{\BBOQ}\APACrefatitle {A Psychometric Analysis of Chess Expertise} {A
  psychometric analysis of chess expertise}.{\BBCQ}
\newblock
\APACjournalVolNumPages{American Journal of Psychology}{118}{}{29-60}.
\PrintBackRefs{\CurrentBib}

\bibitem [\protect \citeauthoryear {%
van~der Linden%
}{%
van~der Linden%
}{%
{\protect \APACyear {2006}}%
}]{%
Vanderlinden:2006}
\APACinsertmetastar {%
Vanderlinden:2006}%
\begin{APACrefauthors}%
van~der Linden, W\BPBI J.%
\end{APACrefauthors}%
\unskip\
\newblock
\APACrefYearMonthDay{2006}{}{}.
\newblock
{\BBOQ}\APACrefatitle {A lognormal model for response times on test items} {A
  lognormal model for response times on test items}.{\BBCQ}
\newblock
\APACjournalVolNumPages{Journal of Educational and Behavioral
  Statistics}{31}{}{181-204}.
\PrintBackRefs{\CurrentBib}

\bibitem [\protect \citeauthoryear {%
van~der Linden%
}{%
van~der Linden%
}{%
{\protect \APACyear {2007}}%
}]{%
vanderLinden:2007}
\APACinsertmetastar {%
vanderLinden:2007}%
\begin{APACrefauthors}%
van~der Linden, W\BPBI J.%
\end{APACrefauthors}%
\unskip\
\newblock
\APACrefYearMonthDay{2007}{}{}.
\newblock
{\BBOQ}\APACrefatitle {A hierarchical framework for modeling speed and accuracy
  on test items} {A hierarchical framework for modeling speed and accuracy on
  test items}.{\BBCQ}
\newblock
\APACjournalVolNumPages{Psychometrika}{72}{}{287–308}.
\PrintBackRefs{\CurrentBib}

\bibitem [\protect \citeauthoryear {%
Van~Zandt%
, Colonius%
\BCBL {}\ \BBA {} Proctor%
}{%
Van~Zandt%
\ \protect \BOthers {.}}{%
{\protect \APACyear {2000}}%
}]{%
vanZandt:2000}
\APACinsertmetastar {%
vanZandt:2000}%
\begin{APACrefauthors}%
Van~Zandt, T.%
, Colonius, H.%
\BCBL {}\ \BBA {} Proctor, R\BPBI W.%
\end{APACrefauthors}%
\unskip\
\newblock
\APACrefYearMonthDay{2000}{}{}.
\newblock
{\BBOQ}\APACrefatitle {A comparison of two response time models applied to
  perceptual matching} {A comparison of two response time models applied to
  perceptual matching}.{\BBCQ}
\newblock
\APACjournalVolNumPages{Psychonomic Bulletin \& Review}{7}{}{208–256}.
\PrintBackRefs{\CurrentBib}

\bibitem [\protect \citeauthoryear {%
Vickers%
}{%
Vickers%
}{%
{\protect \APACyear {1970}}%
}]{%
Vickers:1970}
\APACinsertmetastar {%
Vickers:1970}%
\begin{APACrefauthors}%
Vickers, D.%
\end{APACrefauthors}%
\unskip\
\newblock
\APACrefYearMonthDay{1970}{}{}.
\newblock
{\BBOQ}\APACrefatitle {Evidence for an accumulator model of psychophysical
  discrimination} {Evidence for an accumulator model of psychophysical
  discrimination}.{\BBCQ}
\newblock
\APACjournalVolNumPages{Ergonomics}{13}{}{37-58}.
\PrintBackRefs{\CurrentBib}

\bibitem [\protect \citeauthoryear {%
Walke%
}{%
Walke%
}{%
{\protect \APACyear {2010}}%
}]{%
walke_example_2010}
\APACinsertmetastar {%
walke_example_2010}%
\begin{APACrefauthors}%
Walke, R.%
\end{APACrefauthors}%
\unskip\
\newblock
\APACrefYearMonthDay{2010}{{\APACmonth{05}}}{}.
\newblock
\APACrefbtitle {Example for a Piecewise Constant Hazard Data Simulation in
  {{R}}} {Example for a piecewise constant hazard data simulation in {{R}}}\
  \APACbVolEdTR{\PrintOrdinal{Zeroth}\ \BEd}{\BTR{}\ \BNUM\ TR-2010-003}.
\newblock
\APACaddressInstitution{{Rostock}}{{Max Planck Institute for Demographic
  Research}}.
\newblock
\begin{APACrefDOI} 10.4054/MPIDR-TR-2010-003 \end{APACrefDOI}
\PrintBackRefs{\CurrentBib}

\bibitem [\protect \citeauthoryear {%
Wang%
, Fan%
, Chang%
\BCBL {}\ \BBA {} Douglas%
}{%
Wang%
\ \protect \BOthers {.}}{%
{\protect \APACyear {2013}}%
}]{%
Wang:2013}
\APACinsertmetastar {%
Wang:2013}%
\begin{APACrefauthors}%
Wang, C.%
, Fan, Z.%
, Chang, H.%
\BCBL {}\ \BBA {} Douglas, J.%
\end{APACrefauthors}%
\unskip\
\newblock
\APACrefYearMonthDay{2013}{}{}.
\newblock
{\BBOQ}\APACrefatitle {A semiparametric model for jointly analyzing response
  times and accuracy in computerized testing} {A semiparametric model for
  jointly analyzing response times and accuracy in computerized
  testing}.{\BBCQ}
\newblock
\APACjournalVolNumPages{Journal of Educational and Behavioral
  Statistics}{38(4)}{}{381-417}.
\PrintBackRefs{\CurrentBib}

\end{thebibliography}

\end{document}